\documentclass[conference]{IEEEtran}
\usepackage[hidelinks]{hyperref}
\IEEEoverridecommandlockouts

\usepackage{cite}
\usepackage{amsmath,amssymb,amsfonts}
\usepackage{algorithmic}
\usepackage{graphicx}
\usepackage{textcomp}
\usepackage{xcolor}
\usepackage{verbatim}
\usepackage{booktabs}
\usepackage{hyperref}
\usepackage{subcaption}

\def\BibTeX{{\rm B\kern-.05em{\sc i\kern-.025em b}\kern-.08em
    T\kern-.1667em\lower.7ex\hbox{E}\kern-.125emX}}
\begin{document}

\title{Organizational Security Resource Estimation via Vulnerability Queueing}

\author{
\IEEEauthorblockN{
Abdullah Y.\ Etcibasi,
Zachary Dobos,
C.\ Emre Koksal
}
\IEEEauthorblockA{
Department of Electrical and Computer Engineering\\
The Ohio State University\\
Columbus, OH, USA\\
\{etcibasi.1, dobos.20, koksal.2\}@osu.edu
}
}

\maketitle

\begin{abstract}
We provide an approach that closely estimates an organization’s cyber resources directly from vulnerability timestamps, using a non‑stationary queueing framework. Traditional attack‑surface metrics operate on static snapshots, ignoring the core attack-defense dynamics within information systems, which exhibit a bursty, heavy‑tailed, and capacity‑constrained behavior. Our approach to modeling such dynamics is based on a queueing abstraction of attack surfaces. We utilize a segmentation method to identify piecewise‑stationary regimes via Gaussian‑mixture modeling (GMM) of queue‑length distributions. We fit segment‑specific arrival, service, and resource parameters through the minimization of Kullback-Leibler divergence (KL) between the empirical and estimated distributions. Applied to both large‑scale software supply‑chain data and multi‑year private logistics enterprise cyber‑ticket workflows, the model estimates organizational resources, measured in the time-varying active personnel, and load/rate of output per personnel, solely from bug report and fix timings for software supply chain, and discovery and patch timestamps in the enterprise setting. Our results provide 91-96\% accuracy in estimation of resources, making the dynamic queueing framework an interesting option for understanding attack surface dynamics. Further, our framework exposes resource bottlenecks, establishing a new foundation for predictive workforce planning, patch‑race modeling, and proactive cyber‑risk management.
\end{abstract}


\section{Introduction}
Information systems are continuously exposed to evolving vulnerabilities (e.g software flaws, misconfigurations, emerging zero-day exploits). The set of unpatched vulnerabilities constitutes the time-varying \emph{attack surface}, which is a representative of an organization’s exposure to cyber threats. As infrastructures expand and attackers adopt more automated adaptive techniques, the attack surface evolves dynamically in both size and composition. This evolution is inherently \textit{non-stationary}. Vulnerability discovery patterns fluctuate due to changing development practices and automated scanning tools, while patching capacity varies with staffing and resource allocation. These dynamics often generate heavy backlogs, where vulnerabilities remain unpatched and exploitable. Traditional vulnerability management metrics, such as average open tickets or mean patch time, fail to capture these temporal variations and are therefore insufficient for workforce planning and risk assessment. To address this gap, new analytical tools are required that leverage the full backlog sample path and explicitly account for non-stationary dynamics in both vulnerability arrivals and patching.

Following the attack-surface size as a security metric \cite{manadhata2011attackSurface}, various frameworks  emerged to quantify cyber risk. Existing approaches predominantly treat the snapshots of the attack surface \cite{harry2025countyAttackSurface, jones2011fair, wang2008attackGraphMetric, poolsappasit2012bayesianGraphs}, focusing mostly on long-term averaged risk. Empirical evidence indicates that vulnerability discovery and patching are inherently bursty, heavy-tailed, and subject to regime shifts \cite{haldar2017vulnEpidemics, feutrill2020queueing}. Dynamics suggest that the attack surface is more accurately modeled as a stochastic, capacity-constrained dynamical system. 

This paper addresses these gaps by providing a framework for reconstructing resource dynamics for predictive planning. In particular, we build an empirically validated dynamic queueing framework to model the spatio-temporal evolution of an attack surface. In our approach, we treat vulnerabilities as jobs arriving, patches as services for these jobs, and the queue-length distribution (QLD) to understand the ensemble statistics of the attack surface. The model represents how vulnerabilities arrive, persist, and depart, while accounting for limited defense budgets, attacker–defender asymmetries, and AI-driven acceleration of attack and defense.

We demonstrate that our framework is highly effective in estimating latent organizational parameters, such as effective workforce size and throughput, from event logs. Towards that goal, we focus on two critical security environments: (1) software supply chains, where vulnerabilities propagate via third-party components and package dependencies (open-source data), and (2) large-scale IT service and ticketing systems (proprietary data). We map these traces as parameters within our model (e.g. arrival rates, exploit probabilities). Aligning model dynamics with observed security events, we were able to estimate the actual resources in personnel and their workload with an accuracy of 4-5\%.

In our queueing model for the attack surface, vulnerabilities arrive stochastically and depart via parallel defensive service (patch/mitigation) or exploitation, under a finite aggregate capacity $b$ distributed across $m$ effective servers (abstracted from $G/G/m–b$).  To handle non‑stationarity, we introduce multi‑stage segmentation by fitting a Gaussian-mixture modeling (GMM) to the QLD, aligning mixture components to contiguous time windows, and then identifying segment‑specific $(m,b)$ pairs with arrival/service models (empirically measured to be heavy‑tailed)  via simulation‑based divergence minimization. This reproduces empirical QLDs across regimes and uncovers the effective count personnel handling incidents/alerts $m$ and aggregate throughput capacity $b$. These parameters are inferred from event-level arrival and closure timestamps, enabling cross shift resource allocation and improved staffing forecasting while dynamically varying in time ensuring resources in place to face internal/external factors causing shifts on service backlogs.

We collected extensive amounts of data from an open-source software supply chain (publicly available) and the vulnerability ticketing system of a global supply chain enterprise (private), and use our novel projection techniques to fit the model parameters that are variable in time. We compare with actual values whenever available to demonstrate the accuracy of our resource estimates.

This paper makes the following key contributions:
\begin{itemize}
    \item \textbf{Queueing model of the attack surface:} We build a novel dynamic queueing framework to model the spatio-temporal evolution of an attack surface.
    
    \item \textbf{Segmented Modeling of Non-Stationary Attack Surfaces:} 
    We introduce a multi-stage segmentation approach, based on empirical QLDs, GMM, and targeted simulation-based optimization that reconstructs latent organizational parameters including effective workforce size and aggregate throughput, directly from event logs.
    
    \item \textbf{Validation across two datasets:} We evaluate the framework using software supply chain and from proprietary data of a global supply logistics enterprise's proprietary private dataset, demonstrating high fidelity between empirical and derived QLDs, and close alignment between inferred and independently observed workforce capacity values.
    
    \item \textbf{Organizational Resource Inference from Data:} 
    We demonstrate the fitted queue parameters yield accurate estimates of latent organizational resources, including effective personnel count and service capacity, even when such information is not explicitly available in the data.
\end{itemize}

\section{Related Work}
Early methods offer a general probabilistic decomposition of threats and vulnerabilities,\cite{jones2011fair}, and attack graphs modeled exploit reachability and attacker movement across systems\cite{khosraviFarmad2020bayesian,wang2008attackGraphMetric,manadhata2011attackSurface}. Bayesian and Markov variants extended to conditional dependencies and sequential exploit chains across components, including cloud and Industrial Control Systems environments\cite{ryan2009bayesianThreats,poolsappasit2012bayesianGraphs,zhang2018fuzzyBNics,sabur2022graphBasedCloud,liu2019sequentialAttacks,kotenko2022slrCorrelation}. Recent AI assisted approaches like Prometheus, Graphene, and retrieval augmented LLMs\cite{jin2023prometheus,jin2023graphene,li2024ragAttackGraphs} improve automation but still rely on static system snapshots, assuming immediate patching and overlooking the backlog.

Real‑world vulnerability processing exhibits heavy‑tailed service times (ST), bursty arrivals, and prolonged backlogs. Inter-enterprise environments report patch delays of weeks to months\cite{edgescan2025vulnStats}, long-range dependent vulnerability lifetimes\cite{fang2024llmHack}, and multi‑modal distributions of open vulnerabilities \cite{mei2024arvo}. The ARVO dataset highlights non‑stationarity and multi-regime behavior in vulnerability lifecycles across open‑source ecosystems\cite{mei2024arvo}. These findings necessitate models accounting for finite patching capacity, service heterogeneity, and dynamic backlog evolution. Prior work hints to  queueing‑theoretic interpretations such as analyses of vulnerability‑triage delays\cite{feutrill2020queueing}, these efforts rely on simulated arrivals or single server abstractions, and do not provide data‑calibrated, segment specific queue models capable of reproducing empirical QLDs.

Other studies on dynamic attack–defense interactions use epidemic models\cite{haldar2017vulnEpidemics}, stochastic models~\cite{li2024ragAttackGraphs}, and temporal Bayesian networks\cite{zhang2018fuzzyBNics,sabur2022graphBasedCloud,liu2019sequentialAttacks,kotenko2022slrCorrelation}. These contributions capture temporal dependencies and multi-step exploit patterns but do not model patching as a constrained service, nor incorporate resource bottlenecks that shape real attack surface evolutions. 

Notably, \cite{haldar2017vulnEpidemics} introduces a queue-inspired vulnerability lifecycle model to characterize discovery, patching, and exploitation processes. However, this formulation is embedded within an epidemic framework and does not model the attack surface as a standalone dynamic system, nor does it capture backlog-driven queue length dynamics or data-calibrated vulnerability distributions. To the best of the authors’ knowledge, no prior work provides a fully dynamic, queueing-theoretic model of the attack surface that is calibrated to real-world vulnerability data and capable of reproducing empirical queue length distributions.

These limitations become even more pronounced in large-scale and rapidly evolving ecosystems. Ecosystem-scale studies show inter-organizational correlations, such as shared zero-day shocks and vendor-linked cascades\cite{harry2025countyAttackSurface}, but do not provide tools to quantify backlog propagation or coupled service dynamics. A growing body of work examines AI-amplified cyber risk. Autonomous LLM-based agents capable of API chaining, automated reconnaissance, and exploit generation accelerate both attack frequency and the temporal compression of exploit windows\cite{fang2024llmHack}. Even symmetric AI adoption by defenders and attackers can increase breach rates due to accelerated exploit–patch races, leaving the spatial distribution of backlogs largely unchanged while increasing temporal volatility\cite{berabi2024deepcodeFix}. Benchmarks such as AgentBench\cite{yoran2024assistantbench} and studies on prompt injection\cite{liu2024promptInjection} highlight the fragility of AI agents under adversarial manipulation, but do not provide system-level models linking these behaviors to resource-constrained patching dynamics.

Snapshot‑based risk metrics like probabilistic and graph‑based models\cite{jones2011fair}, general risk‑quantification frameworks, attack graphs \cite{wang2008attackGraphMetric,manadhata2011attackSurface}, Bayesian/Markov variants \cite{ryan2009bayesianThreats,poolsappasit2012bayesianGraphs,khosraviFarmad2020bayesian}, and AI/LLM‑generated attack‑graph systems \cite{jin2023prometheus,jin2023graphene,li2024ragAttackGraphs} summarize the risk in stationary long-term averages. Yet, data from large-scale systems indicate vulnerabilities, patching, and backlogs that are bursty, heavy tailed, and resource limited.  Our model utilizes a queue based race formulation for the exposure, thus capturing vulnerability arrivals, defense and exploit departures, and component dependencies within a dynamic, resource bounded system. This enables organizations to predict resource and capacity based on expected vulnerabilities, setting the foundation of an ERP grade cyber risk management tool for organizations.

\section{System Model and Problem Formulation}

\begin{figure}[!t]
\centering
\includegraphics[width=0.7\linewidth]{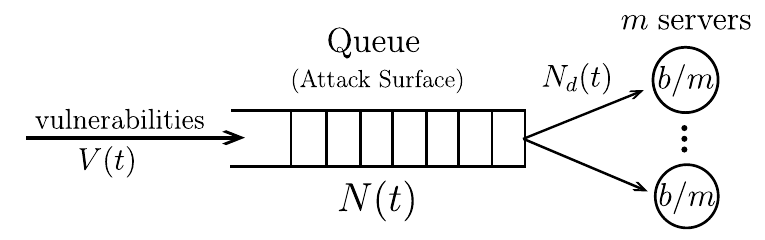}
\caption{Queueing representation of the attack surface.}
\label{fig:model}
\vspace{-1.2em}
\end{figure}

A central contribution of this paper is a framework for modeling the attack surface of a system or organization through a queueing-theoretic approach. Within this model, vulnerabilities are treated as stochastic arrivals that persist in the system until removed through defensive actions. Let $N(t)$ be the instantaneous size of the attack surface, equivalently the queue length, representing the number of active unmitigated vulnerabilities at time $t$. The state evolution of the attack surface is governed by the following discrete-time recursive equation:
\begin{equation}
    N(t+1) = \bigl[ N(t) + V(t) - N_d(t) \bigr]^+ ,
\end{equation}
where $V(t)$ is the number of new vulnerabilities discovered during the interval $[t, t+1)$, $N_d(t)$ represents the number of vulnerabilities mitigated by defensive actions like patching or system reconfigurations. The operator $[x]^+ = \max(x,0)$ enforces non-negativity of the state. This abstraction, shown in Fig.~\ref{fig:model} shows how vulnerabilities $V(t)$ arrive, join the queue of active vulnerabilities $N(t)$, while defenders remove a subset each time step via patching efforts. Since $N(t)$ represents the \emph{attack surface size} at time $t$, larger related to increased exposure and organizational risk. 

In the software supply chain setting (ARVO), $N(t)$ is the reported yet unpatched software bugs, while in the logistics enterprise dataset it represents open security tickets in the organizational workflow. Under this interpretation, standard queueing metrics take on direct operational meaning: the mean queue size reflects expected attack surface size, the waiting-time distribution characterizes how long vulnerabilities persist before resolution;  and tail probability quantifies the likelihood the backlog exceeds a specified risk threshold.

Stochastic processes for the arrival $V(t)$ and the defensive departure $N_d(t)$ are inherently non-stationary. Vulnerability arrivals are bursty and heavy-tailed, while patching times display substantial variability and long-tailed  behavior. Since the departure is capacity-constrained and governed by these service dynamics, both arrival and departure induce temporal heterogeneity and dependence in the backlog $N(t)$. Empirical evidence for these properties is presented in Sections~\ref{sec:Dataset_Overview} and~\ref{sec:Results}.

To capture resource constraints, we adopt a $G/G/m\text{-}b$ formulation, where arrivals and ST follow general distributions and the system has $m$ parallel servers under a finite aggregate resource constraint $b$. Note that our notation differs from the standard Kendall’s framework. While the $G/G/m/k$ model uses $k$ to specify the maximum job capacity \cite{gautam2012analysis}, the proposed $G/G/m\text{-}b$ model uses $m$ as the number of parallel service agents and $b$ as a constraint on total service rate.

In the cyber-security setting considered, $m$ represents the effective number of active patching agents. In the software supply chain case (ARVO), it corresponds to the number of developers or maintainers simultaneously resolving reported vulnerabilities, whereas in the logistics enterprise case it is the number of IT personnel assigned to vulnerability related tickets, with each individual handling at most one ticket at a time. The aggregate capacity parameter $b$ captures the organization’s total patching throughput, measured as completed vulnerabilities or tickets per unit time.

Each server has a fixed service capacity of $b/m$ per unit time, given a total system capacity of $b$. Thus, $m$ captures the scale of the defensive workforce, while $b/m$ captures the average per-agent productivity, reflecting efficiency. This parameterization separates staffing levels from per-person effectiveness, offering a flexible and interpretable framework to analyze attack surface dynamics under resource limits.

This framework naturally extends to incorporate an exploitation process from breaches. In this extended framework, the queue-size evolution becomes $N(t+1) = \bigl[ N(t) + V(t) - (N_d(t) + N_l(t)) \bigr]^+$, where $N_l(t)$ is the number of vulnerabilities exploited during $[t, t+1)$. While our dataset does not include exploitation-specific data, preventing empirical estimation of this process in the present study, the formulation remains fully extendible by introducing additional servers that represent exploitation dynamics.% , and all analytical results presented hold for the extended model. 

Having characterized the attack surface, we next estimate the parameters of the queueing model using a simulation-based optimization framework. Given event-level vulnerability data comprising discovery and patch timestamps, we carefully segment the data into quasi-stationary periods. For each segment, our objective is to identify the parameter set $\theta^*$ that minimizes the statistical distance between the empirical QLD, $\hat{P}(n)$, and the simulated QLD, $P(n,\theta)$, produced by the candidate model. Formally, we solve the optimization problem:

\begin{equation}
    \theta^* = \arg\min_{\theta \in \Theta} \: D_{KL}\left( \hat{P} \, \Vert \, P(\cdot,\theta) \right)
\end{equation}
where $D_{KL}(\cdot \Vert \cdot)$ is the Kullback-Leibler divergence (KL). For two discrete probability mass functions $P(n)$ and $Q(n)$ defined over the same 
support $\mathcal{K}$, the KL is given by
\[
D_{KL}(P \Vert Q)
= \sum_{n \in \mathcal{K}} P(n)\,\log\frac{P(n)}{Q(n)}.
\]

This measure quantifies the discrepancy between the empirical QLD and the distribution generated by a candidate model. The parameter space $\Theta$ is defined for the $G/G/m\text{-}b$ queueing model such that for each segment, $\theta = \{m, b, F_{IA}, F_{ST}\}$, where $m$ is the number of parallel servers, $b$ is the total service capacity, and $F_{IA}$ and $F_{ST}$ are the parametric distributions governing the  inter-arrival (IA) and service time (ST), respectively. 

Note that, our goal is not merely to minimize the KL between a static empirical and simulated QLD. Rather, the KL-based optimization serves as an identification tool for uncovering time-varying resource patterns within a fundamentally non-stationary system. Since vulnerability arrivals and patching dynamics evolve over time, we model the effective workforce size and service capacity as segment-dependent quantities, $m(t)$ and $b(t)$, defined piecewise. By partitioning the observation horizon into quasi‑stationary segments and estimating $(m,b)$ within each, we recover temporal variations in staffing levels and organizational throughput.

The main goal of the framework is to infer how defensive resources evolve over time (captured by $m(t)$ and $b(t)$) so that the resulting queue dynamics reproduce the observed backlog behavior. Although KL gives a rigorous criterion for evaluating statistical fit, the primary goal is the reconstruction of time‑varying organizational capacity from non-stationary data.

\section{A Segmented Framework for Modeling Non-Stationary Attack Surfaces}
\label{sec:Algorithm}
This section presents a data-driven algorithm for fitting a queueing model to non-stationary vulnerability dynamics. Because open-vulnerability trajectories exhibit strong temporal heterogeneity, stationary assumption would not be valid at the global scale. We therefore proceed in four stages: (i) construct the empirical QLD from event-level data, (ii) fit a GMM to identify quasi-stationary regimes, (iii) segment the observation horizon accordingly and estimate segment-specific queue parameters via simulation-based matching, and (iv) validate the model by fitting parametric IA and ST distributions within each segment and comparing the resulting QLDs against empirical and bootstrap baselines.

\subsection{Empirical QLD Construction and Stationarity Assessment}
\label{subsec:QLD_Construction}

The algorithm begins by aligning vulnerability discovery and patching 
timestamps to reconstruct the discrete-time backlog trajectory $N(t)$ 
(e.g., Fig.~\ref{fig:arvo_fullwidth_three}(a) and 
Fig.~\ref{fig:private_fullwidth_three}(a)). 
Rather than analyzing this highly irregular time series directly, 
we construct a time-averaged QLD 
$\hat{P}(n)$ defined as

\begin{equation}
\hat{P}(n)
=
\frac{1}{T}
\sum_{t=1}^{T}
\mathbf{1}_{\{N(t)=n\}},
\end{equation}

where $T$ is the observation horizon and $\mathbf{1}_{\{\cdot\}}$ is the indicator function. Thus, $\hat{P}(n)$ represents the fraction of time the system operates at backlog level $n$.

The time-averaged QLD summarizes the fraction of time the system operates at each backlog level, enabling stationarity assessment and detection of multimodality arising from shifts between low- and high-load regimes. Empirically, it often exhibits multiple modes (e.g., Fig.~\ref{fig:arvo_fullwidth_three}(c) and Fig.~\ref{fig:private_fullwidth_three}(c)), indicating that a single parametric distribution is insufficient. This motivates the use of a Gaussian mixture model (GMM), which provides a flexible yet tractable representation of regime-dependent behavior via a finite mixture of quasi-stationary components.

\subsection{GMM Fitting}
\label{subsec:gmm_fit_qld}

To obtain the quasi-stationary segments, we first fit a univariate GMM with $c$ components to the empirical QLD. Let the continuous mixture density be

\begin{equation}
p_c(x) = \sum_{k=1}^{c} \pi_k \,
\mathcal{N}(x \mid \mu_k, \sigma_k^2),
\end{equation}

where $\pi_k \ge 0$, $\sum_{k=1}^{c} \pi_k = 1$, and 
$\{\mu_k, \sigma_k^2\}_{k=1}^{c}$ are mixture parameters. 
Since the queue length $n$ is discrete, we evaluate the density on 
integer values and normalize to obtain a discrete pmf, \(P_c(n)
=
\frac{p_c(n)}
{\sum_{\ell \in \mathcal{K}} p_c(\ell)},
\quad n \in \mathcal{K}.\)

The parameters are estimated via the Expectation–Maximization (EM) 
algorithm using samples drawn from $\hat{P}$ in proportion to empirical 
time occupancy.

For each mixture order $c$, we evaluate the approximation error using 
the KL divergence $D_{KL}(\hat{P} \,\Vert\, P_c)$. This provides a quantitative measure of fit as a function of mixture 
complexity $c$ (e.g., Fig.~\ref{fig:arvo_fullwidth_three}(b) and 
Fig.~\ref{fig:private_fullwidth_three}(b)).

To select the mixture order, we define the marginal improvement

\begin{equation}
\Delta_c =
D_{KL}(\hat{P} \Vert P_{c-1})
-
D_{KL}(\hat{P} \Vert P_{c}).
\end{equation}

We identify a point of diminishing returns as the smallest $c$ such that $|\Delta_c | < \varepsilon,$ for a predefined tolerance $\varepsilon > 0$. The selected mixture order provides a stable parametric representation of the empirical QLD and serves as a reference benchmark for the subsequent simulation-based identification of the $G/G/m\text{-}b$ parameters 
$\theta = \{m, b, F_{IA}, F_{ST}\}$.

\subsection{Identifying Quasi-Stationary Segments}
\label{subsec:segmentation_gmm}

Let $\{t_0=0 < t_1 < \dots < t_S = T\}$ denote candidate cut points 
on a coarse weekly grid over the observation horizon $[0,T]$. 
Each pair $(t_i,t_j)$ with $i<j$ defines a candidate segment 
$\mathcal{S}_{i,j} = [t_i,t_j)$.

For each candidate segment $\mathcal{S}_{i,j}$ and parameters $(m,b)$, 
we generate a simulated QLD, 
$P^{\mathrm{sim}}_{m,b,i,j}(n)$, using segment-specific IA and ST samples. 
The alignment between the simulated distribution and a selected 
GMM component $P_c$ is evaluated via

\begin{equation}
\mathcal{L}_{i,j}(m,b)
=
D_{KL}\!\left(
P^{\mathrm{sim}}_{m,b,i,j}
\,\Vert\,
P_c
\right).
\end{equation}

For each segment $\mathcal{S}_{i,j}$, we compute provisional 
parameters $(m',b')$ via

\begin{equation}
(m',b')
=
\arg\min_{m,b}
\mathcal{L}_{i,j}(m,b),
\end{equation}
and define the segment score \( \mathcal{L}_{i,j} = \mathcal{L}_{i,j}(m',b').\) Importantly, the objective of this step is not to precisely identify the optimal $(m,b)$ parameters, but rather to evaluate how well a candidate time interval behaves as a quasi-stationary regime. Accordingly, the search over $(m,b)$ is performed on a coarse grid within prespecified ranges to provide a computationally efficient segment-level score. Refined parameter estimates $(m^*,b^*)$ are obtained in the subsequent estimation stage once the segmentation is fixed.

The simulated distribution $P^{\mathrm{sim}}_{m,b,i,j}(n)$ is obtained via bootstrap simulation: IA and ST samples are resampled with replacement from empirical segment data. ST samples are scaled by $m/b$ to satisfy the total capacity constraint, and the resulting queue trajectory is simulated to produce an empirical pmf. Repetitions are averaged until convergence of the running pmf estimate.

Among all feasible segments, we select the mutually exclusive partition identified via coarse grid search that achieves the lowest aggregate score $\sum \mathcal{L}_{i,j}$ over the observation horizon. These segments define quasi-stationary intervals as illustrated by the vertical dotted lines in Fig.~\ref{fig:arvo_fullwidth_three}(a) and Fig.~\ref{fig:private_fullwidth_three}(a).

\subsection{Parametric Fitting and Validation}
\label{subsec:Parameter_Fitting}

Conditioned on the segmentation in 
Section~\ref{subsec:segmentation_gmm}, we refine the queue parameters 
within each segment via a fine-grained search over $(m,b)$. 
For each segment $\mathcal{S}_{i,j}$, we solve

\begin{equation}
(m^*,b^*)
=
\arg\min_{m,b}
D_{KL}\!\left(
P^{\mathrm{sim}}_{m,b,i,j}
\,\Vert\,
P_c
\right),
\end{equation}

using a higher-resolution search around the provisional 
$(m',b')$. The resulting $(m^*,b^*)$ defines the refined service 
configuration.

Given fixed segmentation and $(m^*,b^*)$, we next model IA and ST 
distributions parametrically. For each segment, IA and ST samples 
are extracted from event timestamps and modeled independently.
fWe consider candidate families capable of capturing both light- and 
heavy-tailed behavior, including lognormal, log-logistic, generalized 
Pareto, and generalized extreme-value distributions, as well as 
lognormal GMMs and two-component mixture models fitted via EM to 
capture multi-scale or burst-driven dynamics. The best-performing 
model is selected based on empirical fit (see 
Fig.~\ref{fig:arvo_segment_validation_full}(a) and 
Fig.~\ref{fig:private_four_row}(a)–(b)).

For validation, each segment’s selected parametric model is simulated 
under $(m^*,b^*)$. The resulting segment-level QLDs are aggregated 
according to the GMM weights,

\begin{equation}
P^{\mathrm{agg}}(n)
=
\sum_{s} \pi_s \,
P^{\mathrm{sim}}_{s}(n).
\end{equation}

An aggregate bootstrap QLD is constructed in parallel. The aggregated 
parametric and bootstrap QLDs are compared against the empirical QLD 
and the GMM-implied mixture distribution to assess whether the fitted 
models and segmentation jointly reproduce the observed queue dynamics.

\section{Dataset Overview}
\label{sec:Dataset_Overview}
This section describes the datasets used to validate our queueing-theoretic framework: a large-scale public vulnerability dataset (ARVO) and a multi-year logistics enterprise dataset. ARVO captures reproducible vulnerabilities discovered via large-scale fuzzing, while the enterprise dataset reflects real-world patching dynamics. In both cases, event-level timestamps allow reconstruction of arrival and service processes, enabling modeling of vulnerability disclosures and patch completions as stochastic queueing systems.

\subsection{Software Supply Chain}

The first dataset used to validate our model is ARVO ~\cite{mei2024arvo}, a large-scale, curated collection of reproducible vulnerabilities discovered in open-source C/C++ projects through Google’s OSS-Fuzz infrastructure. ARVO contains over 4{,}000 vulnerability instances spanning multiple years (2017--2024), each accompanied by detailed event-level metadata.

For each vulnerability, the dataset records the \textit{report timestamp} (disclosure or discovery date), the \textit{fix timestamp} (patch date), the \textit{severity level} (low, medium, high, or unknown), the \textit{sanitizer type} that exposed the bug (ASan, MSan, or UBSan), and the \textit{crash category} (e.g., heap-buffer-overflow or use-after-free). These attributes enable precise reconstruction of both arrival and service, allowing us to model vulnerability disclosures as arrivals and patch completions as service events within a queueing-theoretic framework. The dataset spans eight years (2017--2024), covering periods of rapid vulnerability discovery, structured patching efforts, and before stabilization phases. Vulnerability reporting exhibits pronounced burstiness, characterized by sharp spikes and irregular waves of disclosures rather than steady arrivals. 

\begin{figure}[t]
\centering

\begin{subfigure}{0.55\columnwidth}
    \centering
    \includegraphics[width=\linewidth]{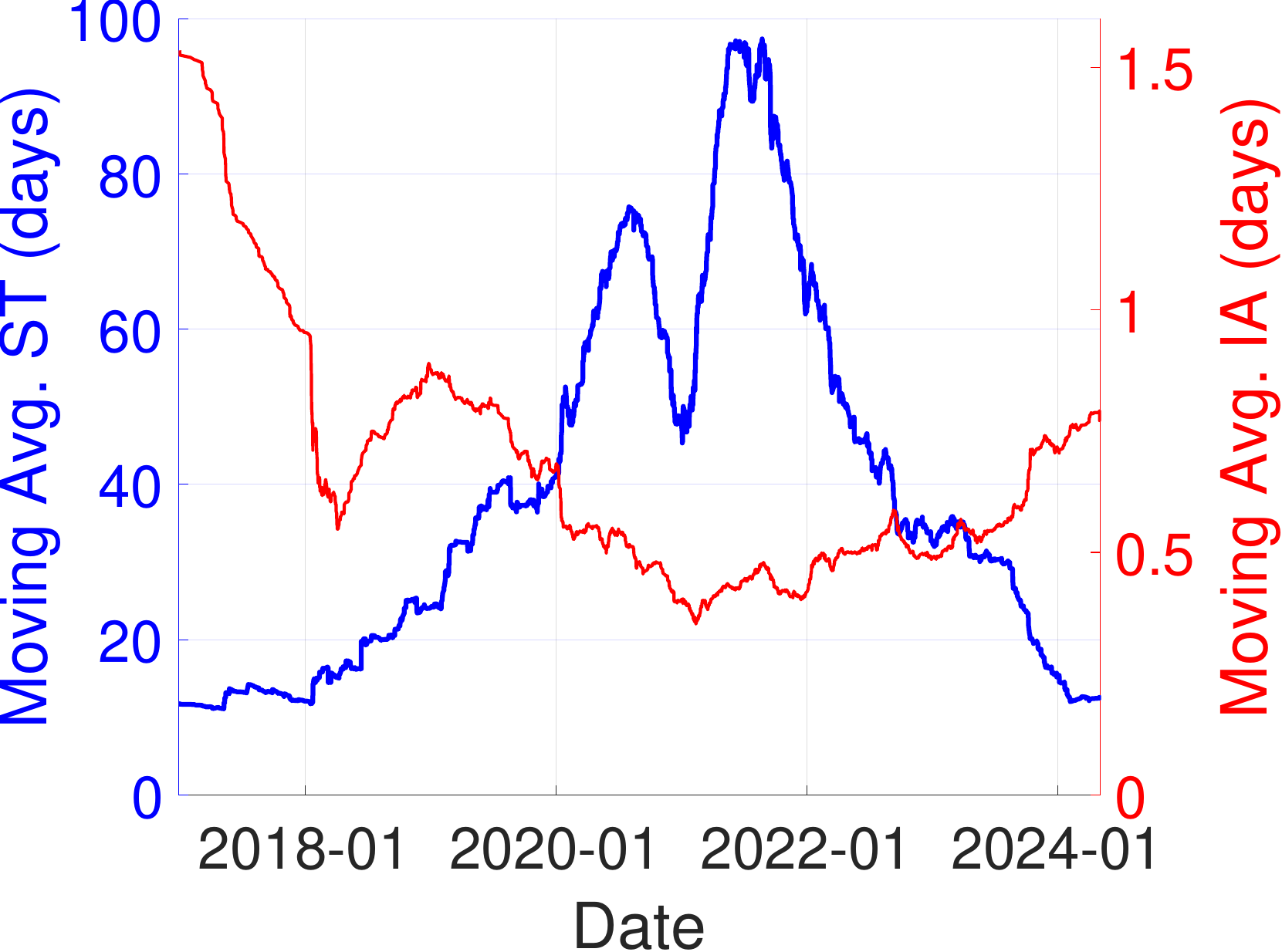}
    \caption{Moving-average IA and ST (500-event window) for ARVO.}
\end{subfigure}
\hfill
\begin{subfigure}{0.40\columnwidth}
    \centering
    \includegraphics[width=\linewidth]{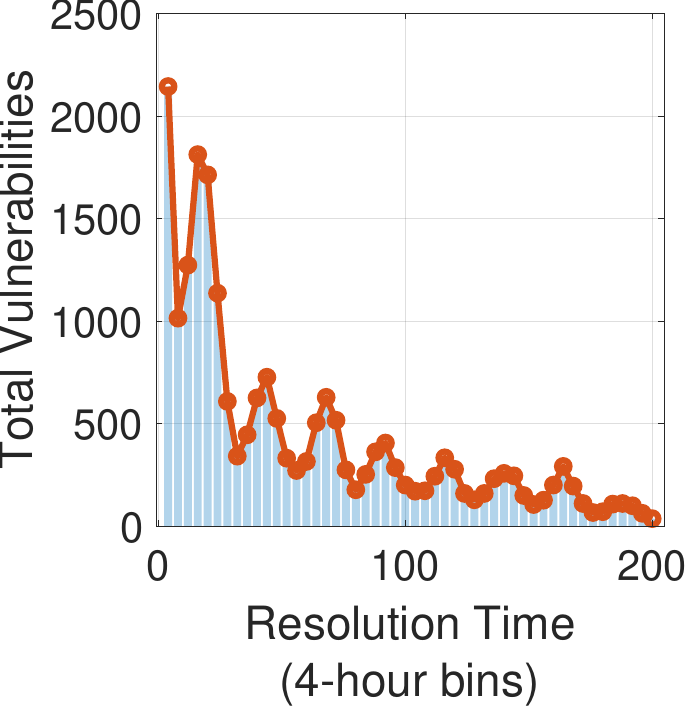}
    \caption{Resolution-time distribution (4-hour bins) for the logistics enterprise.}
\end{subfigure}

\caption{Temporal dynamics in the ARVO and logistics enterprise datasets.}
\label{fig:temporal_dynamics_singlecol}
\vspace{-1.2em}
\end{figure}

The dataset exhibits several properties central to queuing analysis. Both IA and ST exhibit pronounced heavy-tailed behavior. Light-tailed models such as the exponential distribution systematically underestimate the empirical tails, whereas lognormal and mixture models provide substantially improved fits (see Fig.~\ref{fig:arvo_segment_validation_full}(a)). The distributions extend over long durations, with non-negligible mass in the tail, indicating clustered arrivals and prolonged exposure windows.

Second, the processes are clearly non-stationary. Fig.~\ref{fig:temporal_dynamics_singlecol}(a) shows the moving averages of IA and ST computed over a 500-event window. Both series fluctuate substantially over time rather than stabilizing around constant levels. In particular, ST exhibit pronounced growth and contraction phases, while IA times vary across distinct temporal regimes. These sustained temporal variations confirm the arrival and service cannot be adequately described by a single stationary model.

\begin{figure*}[t]
\centering

\begin{subfigure}{0.32\textwidth}
    \centering
    \includegraphics[width=\linewidth]{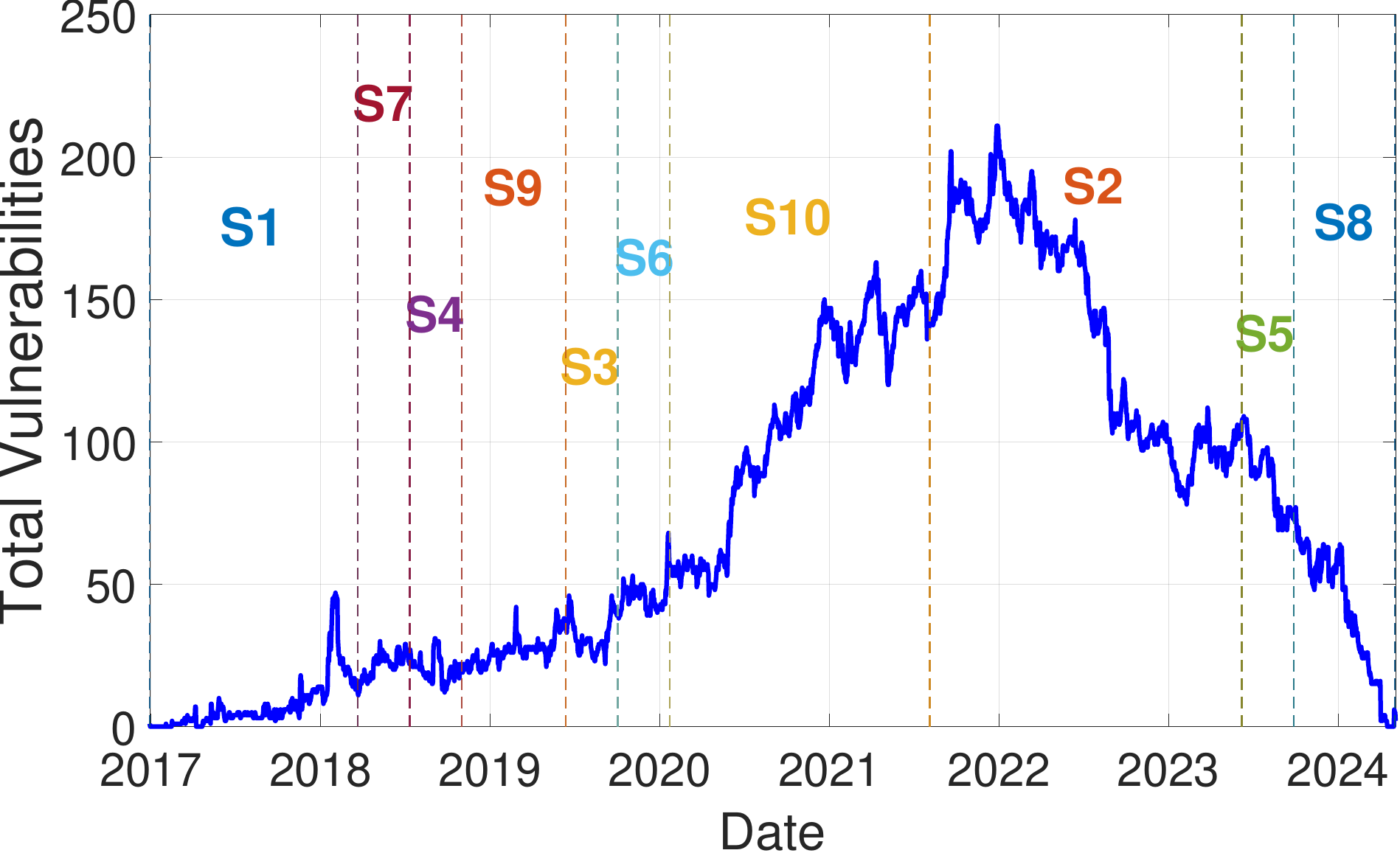}
    \caption{Queue-length evolution with cut points defining quasi-stationary segments.}
\end{subfigure}
\hfill
\begin{subfigure}{0.22\textwidth}
    \centering
    \includegraphics[width=\linewidth]{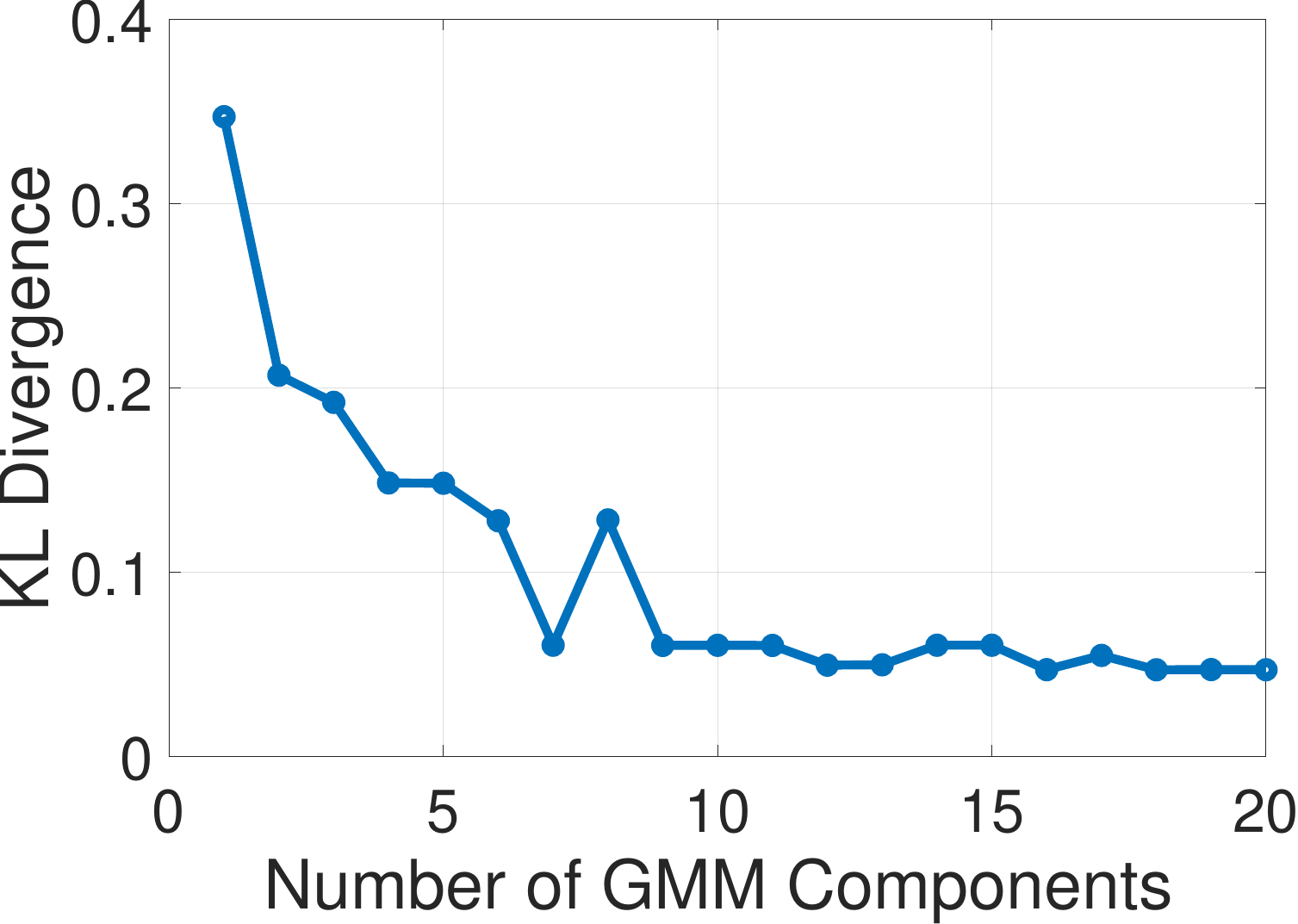}
    \caption{KL divergence between the empirical QLD and GMMs vs. number of components.}
\end{subfigure}
\hfill
\begin{subfigure}{0.38\textwidth}
    \centering
    \includegraphics[width=\linewidth]{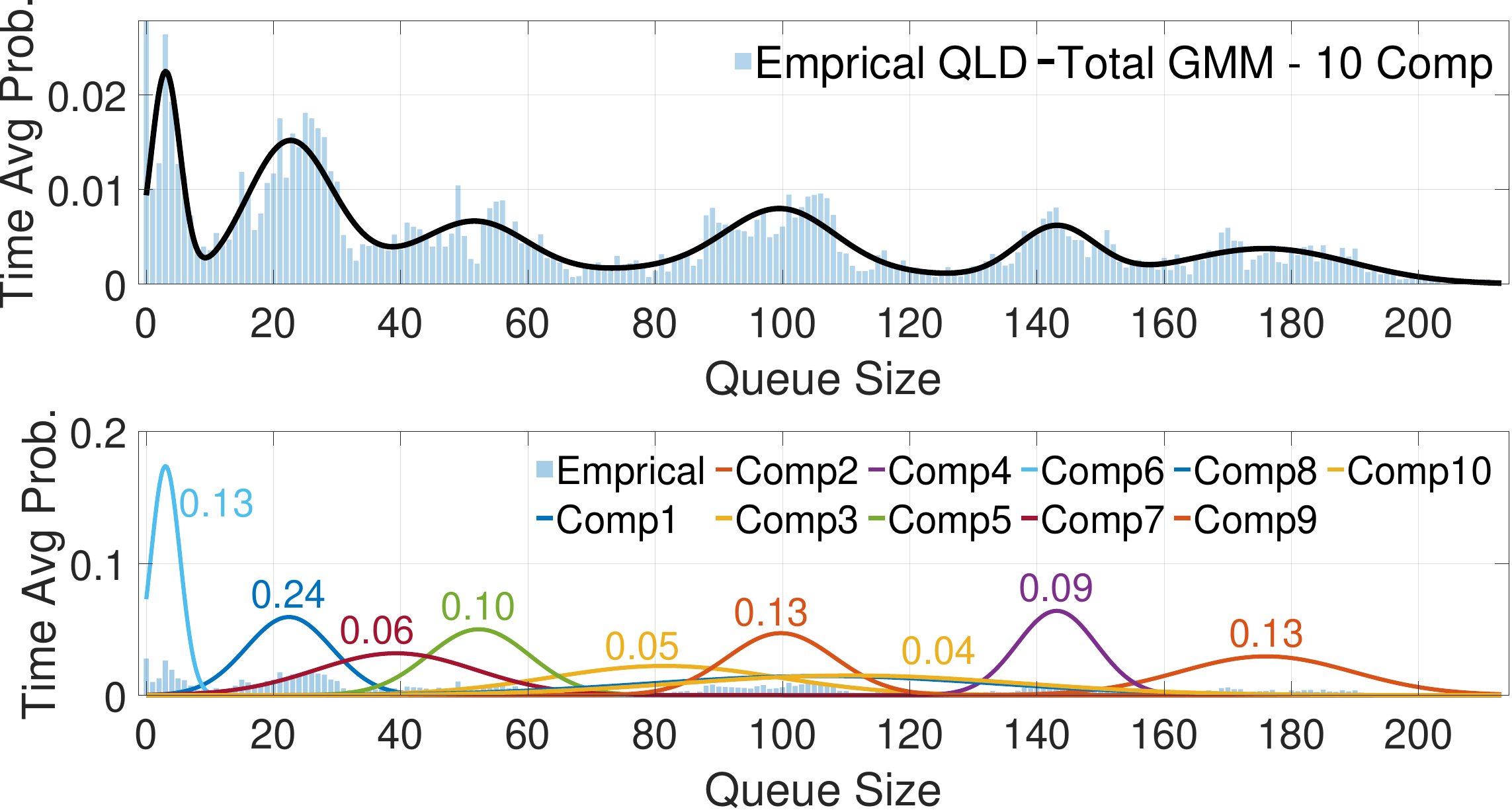}
    \caption{Ten-component GMM fit to the empirical QLD with individual components and aggregate mixture.}
\end{subfigure}

\caption{Segmentation and mixture-model fitting results for the ARVO dataset.}
\label{fig:arvo_fullwidth_three}
\vspace{-1.2em}
\end{figure*}

The heavy-tailed behavior and pronounced temporal non-stationarity provide strong empirical justification for adopting a segmented $G/G/m-b$ modeling framework. Prior vulnerability datasets only aggregate disclosure statistics while ARVO enables a full stochastic  characterization of vulnerability arrivals and lifetimes. This event-level granularity makes it well-suited for queueing-theoretic modeling of the attack surface.

\subsection{Logistics Enterprise}
The second dataset used to validate our model originates from a large logistics enterprise. It consists of a multi‑year organizational cyber-security dataset derived from the enterprise’s vulnerability and ticketing workflow. The data were extracted from their primary source systems and subsequently cleansed. Each record represents a human or system generated ticket or alert. The resulting time series exhibits pronounced burstiness, characterized by sharp spikes and irregular waves of disclosures rather than smooth or stationary arrivals.

For each ticket or alert, the dataset includes the timestamp at which it was opened (discovery, report, or creation), the user assigned \textit{severity level} (low, medium, high, or unknown), the incident category, the \textit{resolver} (service desk or work group), and the resolution timestamp. We selected the resolution timestamp rather than the final close or informed dates because it most accurately reflects when the fix was implemented. Close dates frequently include workflow buffers of up to two weeks to allow time for end user confirmation, and they may not reflect the true time of patching. If a user reopens or provides feedback on an incident, the resolution timestamp is adjusted accordingly. As part of date cleansing, all tickets still in progress without a valid resolution timestamp were removed.
These preprocessing steps enabled reconstruction of the weekly attack‑surface trajectory $N(t)$. To avoid boundary artifacts, we restrict the dataset to 2019–2025. 

Fig.~\ref{fig:private_fullwidth_three}(a) shows three notable regime shifts within this interval. From 2019 to early 2020, the baseline number of weekly incidents remains near 200, interrupted by a sharp spike exceeding 400 during a week in 2019. This spike aligns with an organizational system implementation that caused widespread BitLocker Recovery Key issues, prompting an unusually high volume of telephone submitted tickets to the cyber‑security help desk. Early 2020 begins with another spike above 300, attributable to the organization’s rapid transition to remote work during the onset of COVID‑19. Afterward, the baseline drops to ~100, with minor annual spikes until 2025, when the series shifts dramatically upward toward 500. We attribute this rise to the introduction of new, highly sensitive endpoint detection mechanisms that generate more alerts.

Fig.~\ref{fig:temporal_dynamics_singlecol}(b) visualizes the distribution of vulnerabilities binned into 4 hour resolution intervals. To highlight the structural pattern, we filter the data to include only tickets resolved between 4 and 200 hours. Intervals below 4 hours are omitted because their large counts obscure the characteristic 24 hour work‑cycle pattern. Intervals above 200 hours are excluded due to extreme long‑tail outliers. The resulting distribution shows a consistent 24 hour wave across all resolution times and an exponential decay in total vulnerabilities as resolution time increases. We interpret this pattern as reflecting the combined influence of system behavior, service level agreements, and human work‑hour dynamics on patching.

\section{Results}
\label{sec:Results}
This section presents empirical validation of the proposed segmentation-based queueing framework across two distinct datasets. Beyond fitting QLDs, our central objective is to infer the system parameters and assess whether these inferred $(m,b)$ values are consistent with actual capacity. We first evaluate the methodology on the ARVO vulnerability dataset to assess its ability to capture non-stationary and heavy-tailed queue dynamics. We then apply the same framework to the logistics enterprise ticketing dataset, where the inferred $(m,b)$ parameters closely align with independently observed workforce size and throughput data, providing strong external validation of the model.

\begin{figure*}[t]
\centering

\begin{subfigure}{0.28\textwidth}
    \centering
    \includegraphics[width=\linewidth]{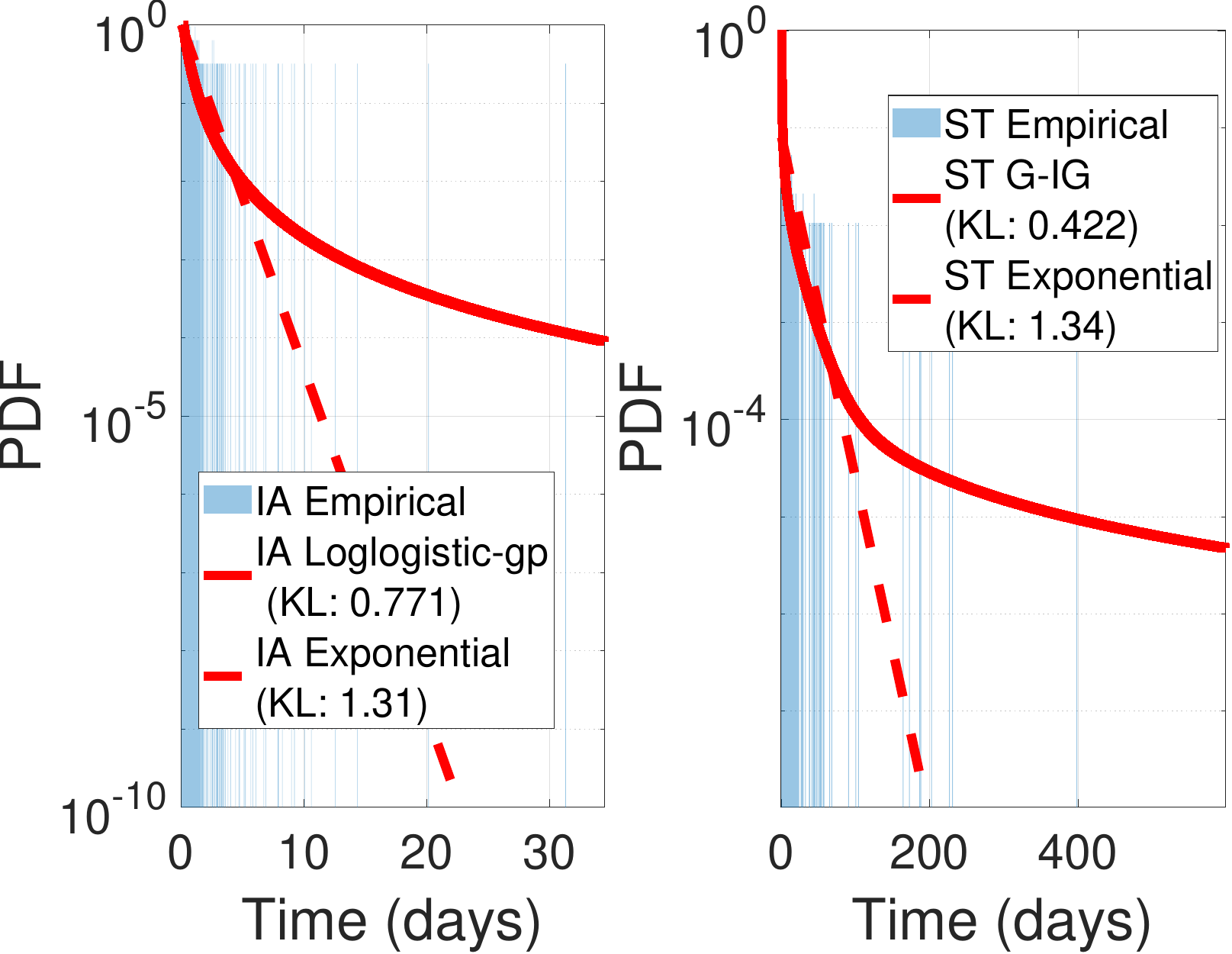}
    \caption{Best-fitting IA and ST models for Component 1 (Weeks 0–64). Heavy-tailed mixtures achieve the lowest KL.}
\end{subfigure}
\hfill
\begin{subfigure}{0.18\textwidth}
    \centering
    \includegraphics[width=\linewidth]{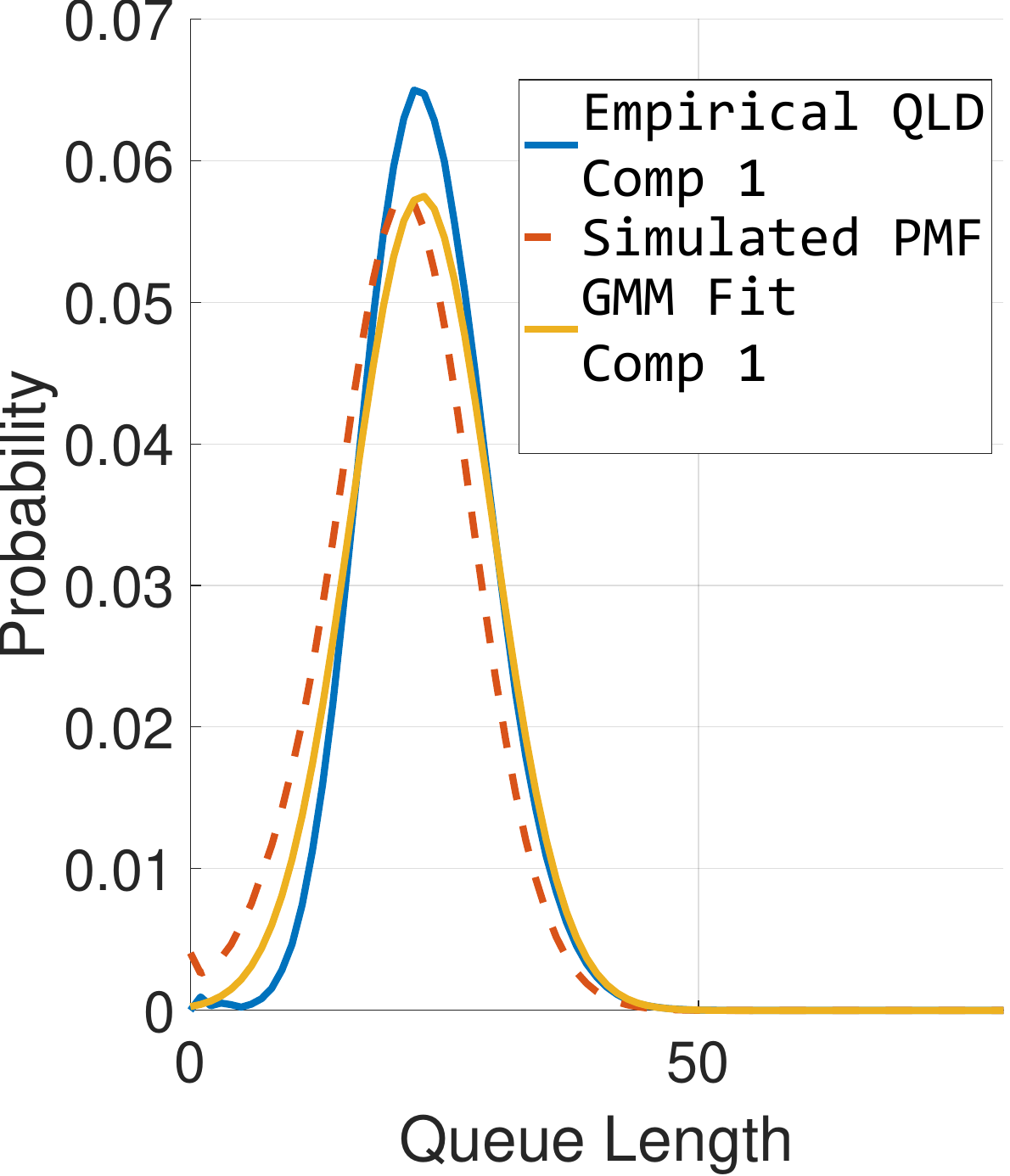}
    \caption{Simulated vs. empirical QLD for Component 1 under fitted parameters.}
\end{subfigure}
\hfill
\begin{subfigure}{0.48\textwidth}
    \centering
    \includegraphics[width=\linewidth]{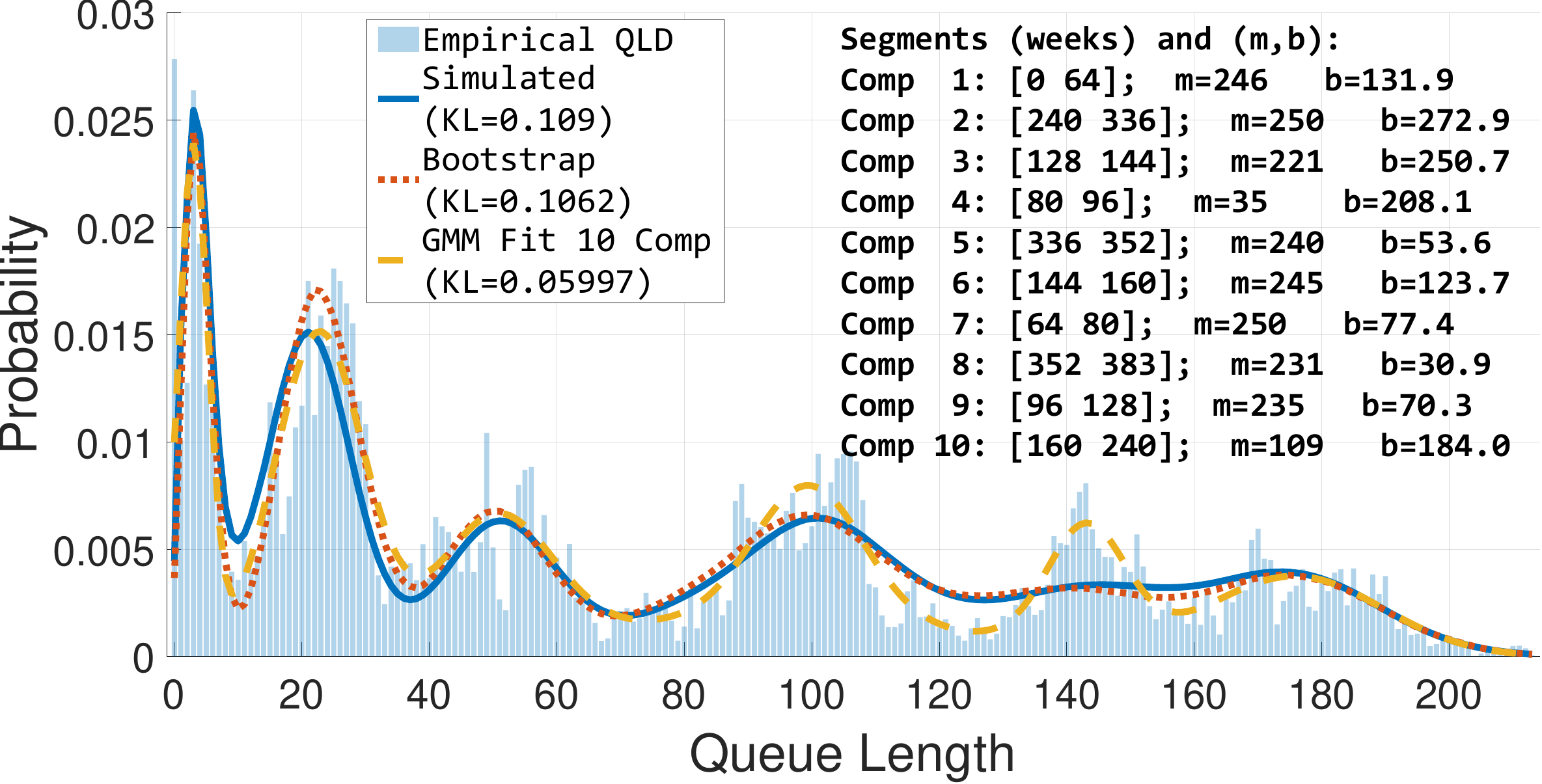}
    \caption{Overall empirical QLD with segmented $G/G/m\text{-}b$ simulation, bootstrap estimate, and 10-component GMM benchmark.}
\end{subfigure}

\caption{Segment-level fitting and overall QLD validation for ARVO. The segmented $G/G/m\text{-}b$ model closely matches empirical behavior (KL 0.109).}
\label{fig:arvo_segment_validation_full}
\vspace{-1em}
\end{figure*}

\subsection{Software Supply Chain}

We begin by applying the segmentation-based queue modeling framework of Section~\ref{sec:Algorithm} to ARVO. The objective is to determine whether the segmented $G/G/m-b$ abstraction can faithfully reproduce the empirical attack surface dynamics observed in ARVO.

First we reconstruct the attack surface trajectory $N(t)$ from vulnerability report (arrival) and fix (service completion) timestamps. The resulting queue-length evolution over time is shown in Fig.~\ref{fig:arvo_fullwidth_three}(a). The system exhibits pronounced non-stationarity. Periods of backlog growth are followed by intervals of rapid reduction, indicating time-varying imbalance between vulnerability discovery and patching.

From this trajectory, we construct the empirical QLD (Fig.~\ref{fig:arvo_fullwidth_three}(c)). The distribution is clearly multi-modal, with multiple distinct peaks corresponding to different regimes. This observation confirms single stationary queue model is unlikely to capture the full dynamics of the attack surface.

Following Section~\ref{subsec:gmm_fit_qld}, we approximate the empirical QLD using GMMs with varying numbers of components. The KL as a function of mixture complexity is shown in Fig.~\ref{fig:arvo_fullwidth_three}(b). Fit quality improves sharply up to 10-components and then exhibits diminishing returns. Based on this elbow behavior, we select a 10-component GMM to represent distinct quasi-stationary regimes.

Each mixture component is then mapped back onto the time axis by identifying contiguous intervals whose segment-level QLDs most closely match the corresponding GMM component in terms of KL. The resulting segmentation is shown in Fig.~\ref{fig:arvo_fullwidth_three}(a), where the identified cut points partition the queue-length trajectory into quasi-stationary regimes. Each segment corresponds to a distinct condition of the attack surface, characterized by its own arrival and service dynamics. The clear structural shifts in $N(t)$ across the marked boundaries visually validate the distributional segmentation obtained from the GMM-based procedure.

Within each identified segment, we extract IA and ST samples and fit parametric mixtures, as described in Section~3.4. 

Fig.~\ref{fig:arvo_fullwidth_three}(a) shows representative IA and ST fits for an early segment (Weeks 0--64). The best-fitting models are a log logistic–generalized Pareto mixture for arrivals (KL = 0.771) and a gamma–inverse Gaussian (G-IG) for ST (KL = 0.422). The resulting simulated QLD, shown in Fig.~\ref{fig:arvo_segment_validation_full}(b), closely tracks the empirical distribution for that segment.

These results confirm that, once segmented, vulnerability arrivals and patching times for intervals admit stable \textbf{heavy-tailed} distributions that yield accurate queue-length predictions.

Finally, we aggregate the segment-wise simulations into a global model of the attack surface. Fig.~\ref{fig:arvo_segment_validation_full}(c) compares the empirical QLD with three baselines: a bootstrap simulation constructed from the empirical IA and ST, a 10-component GMM approximation, and the segmented $G/G/m-b$ model.

The segmented model achieves a divergence of $0.1074$, closely matching the bootstrap estimate ($0.1069$) and approaching the GMM benchmark ($0.05997$). Across segments, the number of servers $m$ remains stable (typically 221--250), while the effective capacity $b$ varies widely (30.9 to 272.9), reflecting substantial temporal fluctuations in patching throughput. These variations correspond to changes in patching intensity over time and shows how defensive resource allocation directly shapes the attack surface distribution.

The ARVO results reveal several important quantitative insights. First, the vulnerability arrival and patching exhibit pronounced non-stationarity and heavy-tailed behavior, as evidenced by the fitted IA and ST mixtures and the clear regime shifts visible in Fig.~\ref{fig:arvo_fullwidth_three}(a). Second, the identified quasi-stationary segments produce a global simulated QLD with KL $0.109$, closely matching the empirical bootstrap benchmark ($0.1062$). This close quantitative agreement with the empirical QLD demonstrates that the segmented $G/G/m\text{-}b$ model captures the essential structural dynamics of the backlog.

The recovered parameters reveal meaningful temporal variation in defensive capacity. While the effective number of active servers $m$ remains relatively stable across regimes (typically in the range $221$--$250$), the aggregate service capacity $b$ varies significantly (~$30.9$ to $272.9$), reflecting substantial shifts in patching intensity over time. These variations are consistent with observable phases of backlog expansion and contraction in Fig.~\ref{fig:arvo_fullwidth_three}(a), demonstrating that the model does not merely reproduce distributional features but also identifies interpretable changes in throughput.

These results provide strong empirical validation that the proposed non-stationary queue reconstruction framework accurately captures the attack surface evolution observed in ARVO. In the next section, we apply the same methodology to the logistics enterprise dataset, where independent workforce and throughput records enable direct quantitative validation of the inferred $(m,b)$ parameters.

\begin{figure*}[t]
\centering

\begin{subfigure}{0.26\textwidth}
    \centering
    \includegraphics[width=\linewidth]{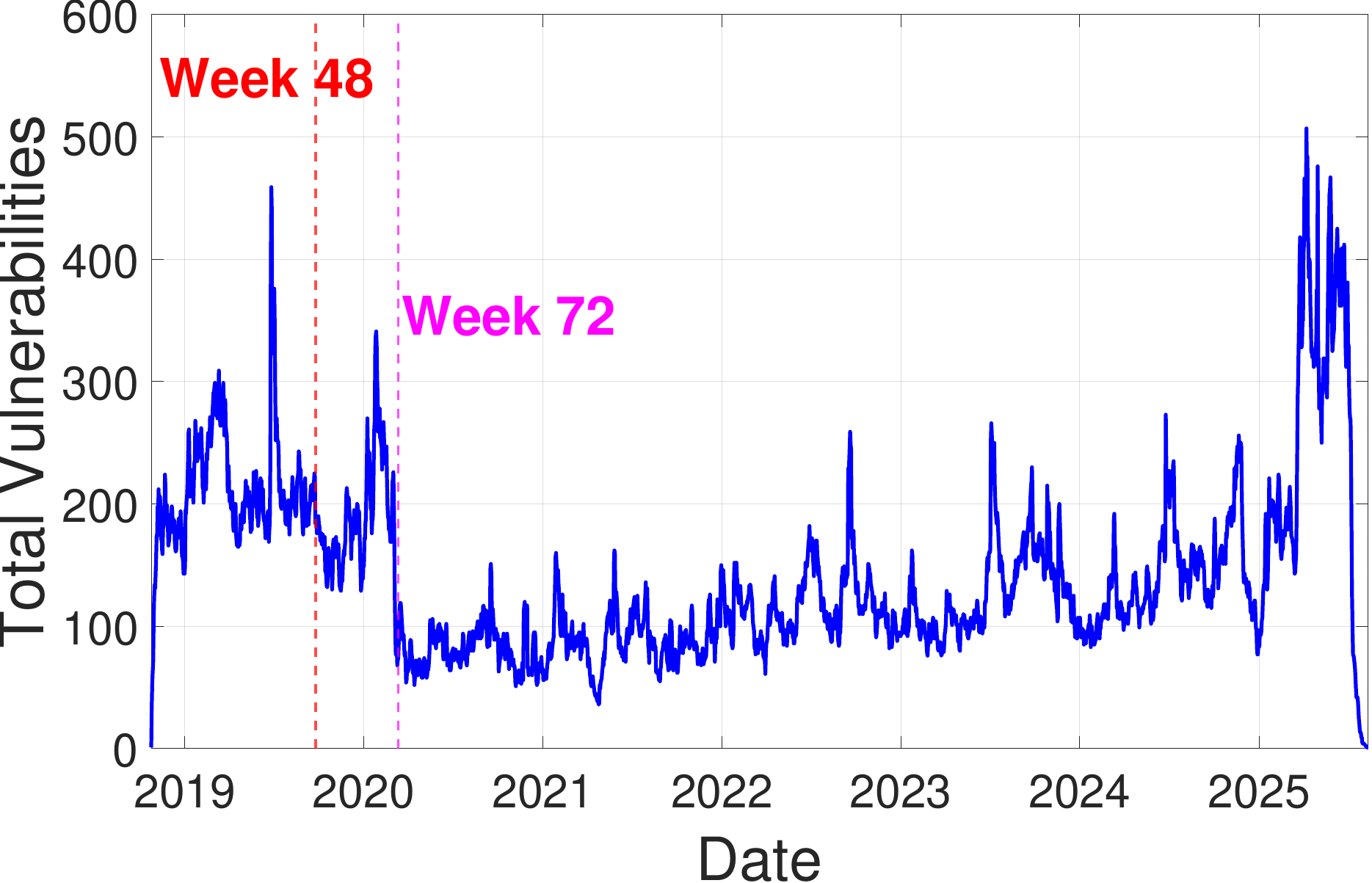}
    \caption{Queue-length evolution with cut points defining quasi-stationary segments.}
\end{subfigure}
\hfill
\begin{subfigure}{0.24\textwidth}
    \centering
    \includegraphics[width=\linewidth]{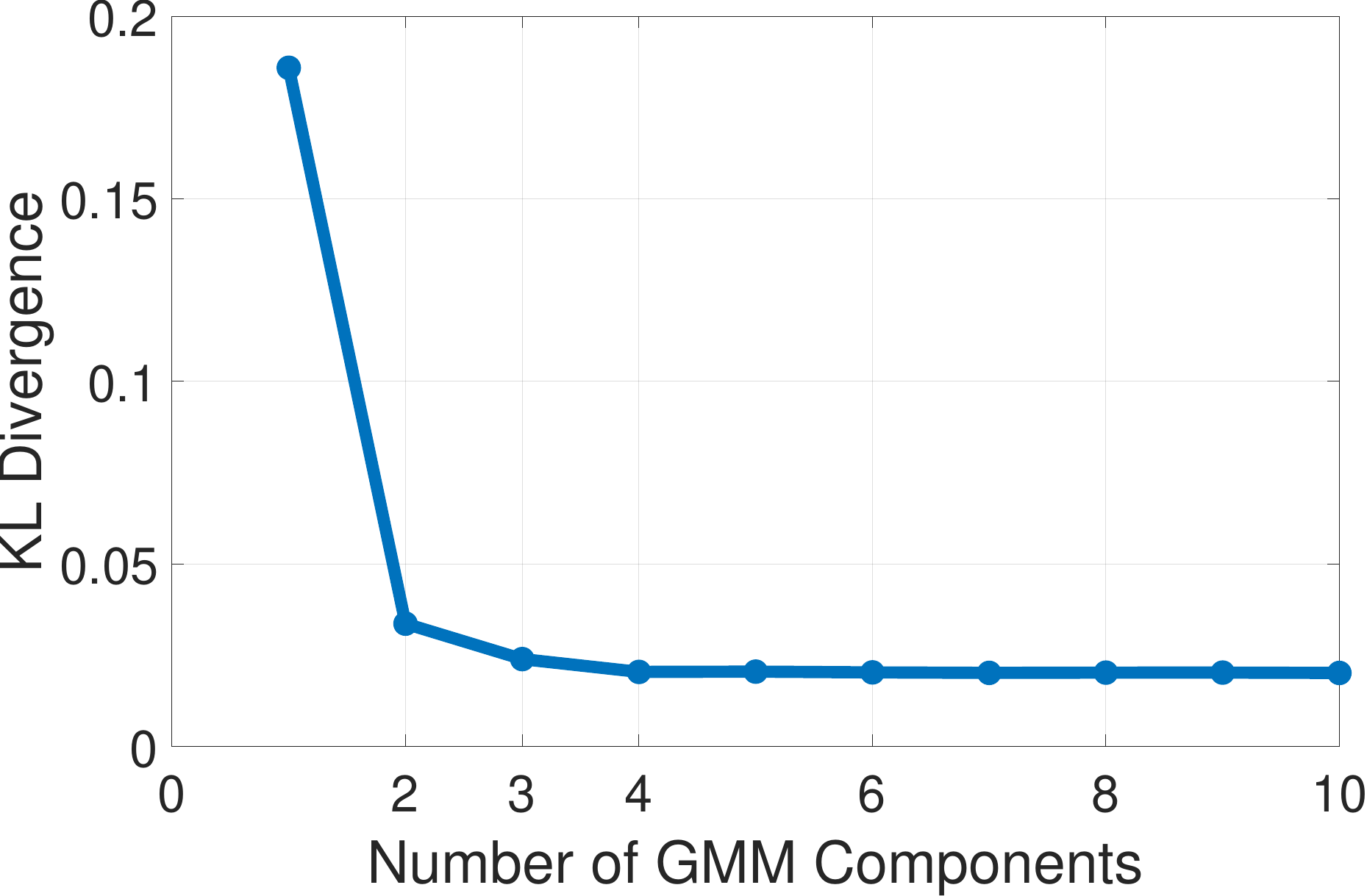}
    \caption{KL divergence between the empirical QLD and GMMs vs. number of components.}
\end{subfigure}
\hfill
\begin{subfigure}{0.46\textwidth}
    \centering
    \includegraphics[width=\linewidth]{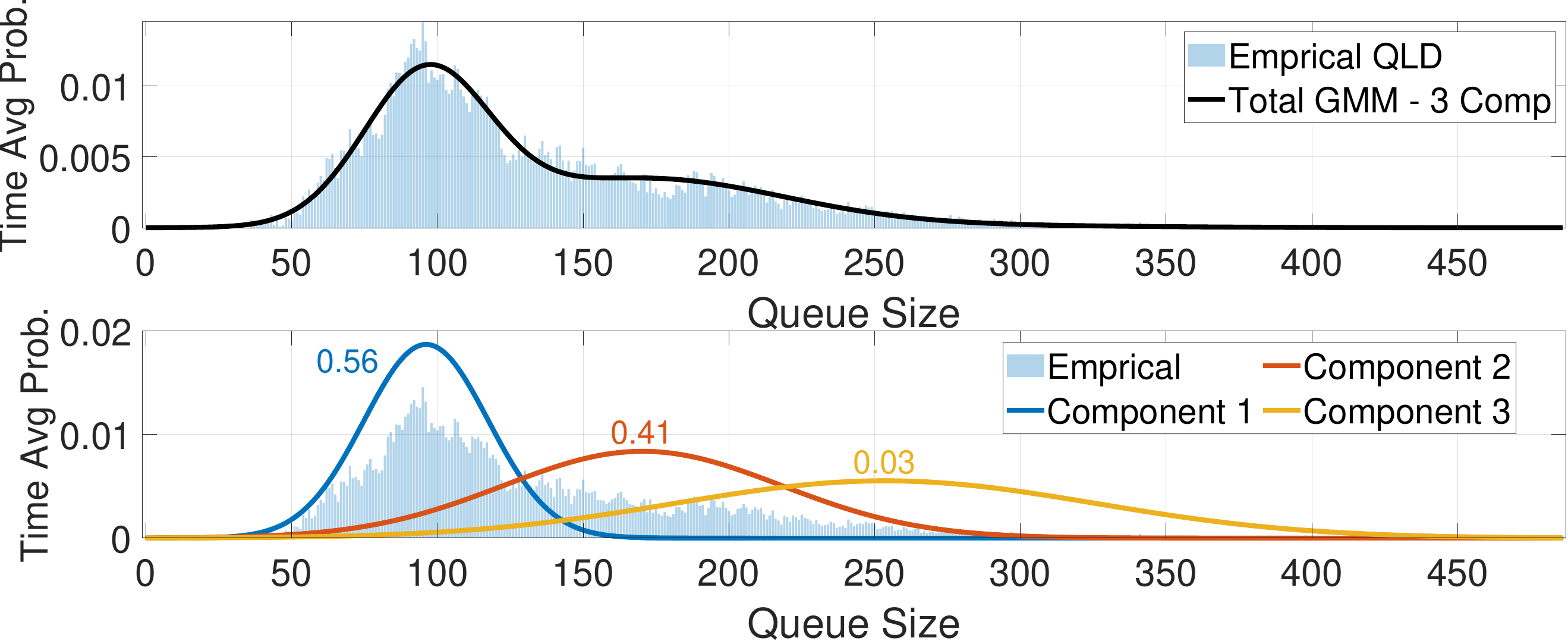}
    \caption{Three-component GMM fit to the empirical QLD with individual components and aggregate mixture.}
\end{subfigure}

\caption{Segmentation and mixture-model fitting results for the logistics enterprise dataset.}
\label{fig:private_fullwidth_three}
\vspace{-1em}
\end{figure*}

\subsection{Logistics Enterprise}
As described in Section~\ref{subsec:QLD_Construction}, we begin by reconstructing the attack surface queue from vulnerability discovery (arrival) and patching (departure) timestamps. Fig.~\ref{fig:private_fullwidth_three}(a) depicts the resulting queue-length trajectory over time. The system exhibits pronounced non-stationarity. During the initial period (2019-2020), the queue remains persistently large, indicating sustained overload or insufficient defensive capacity, after which the backlog decreases sharply and stabilizes at a significantly lower level. A second transient surge is observed toward the end of the observation horizon (after 2025). To focus the analysis on representative regimes and exclude these boundary effects, we restrict attention to the interval spanning 2019–2025 by cropping the data accordingly. 

Using the cropped queue-length trajectory, we construct the empirical QLD (Fig.~\ref{fig:private_fullwidth_three}(c)). The distribution confirms the presence of non-stationarity and the empirical QLD multi-modal, exhibiting a dominant peak around a QLD of ~100 with a substantial probability mass concentrated near 150. This structure indicates that the system operates under multiple distinct load conditions rather than fluctuating around a single stationary regime. Motivated by this observation, we employ GMMs to identify and separate these quasi-stationary regimes within the data.

In Fig.~\ref{fig:private_fullwidth_three}(b), we report the KL between the empirical QLD and GMMs with varying numbers of components. The marginal improvement in KL diminishes sharply beyond three components, indicating additional mixture components provide limited explanatory benefit. Accordingly, we adopt a three-component GMM, corresponding to three distinct quasi-stationary regimes. Notably, this contrasts with ARVO, where 10-components were required to capture the pronounced temporal variability visible in the queue-size-versus-time trajectories. In comparison, the organizational dataset exhibits substantially more stable dynamics, consistent with our prior expectation that structured enterprise workflows would produce fewer regime shifts. 

The individual GMM components and their weighted fit are shown in Fig.~\ref{fig:private_fullwidth_three}(c), where the three-component mixture closely matches the empirical QLD. With the number of regimes identified.

\begin{figure*}[t]
\centering

\begin{subfigure}{0.16\textwidth}
    \centering
    \includegraphics[width=\linewidth]{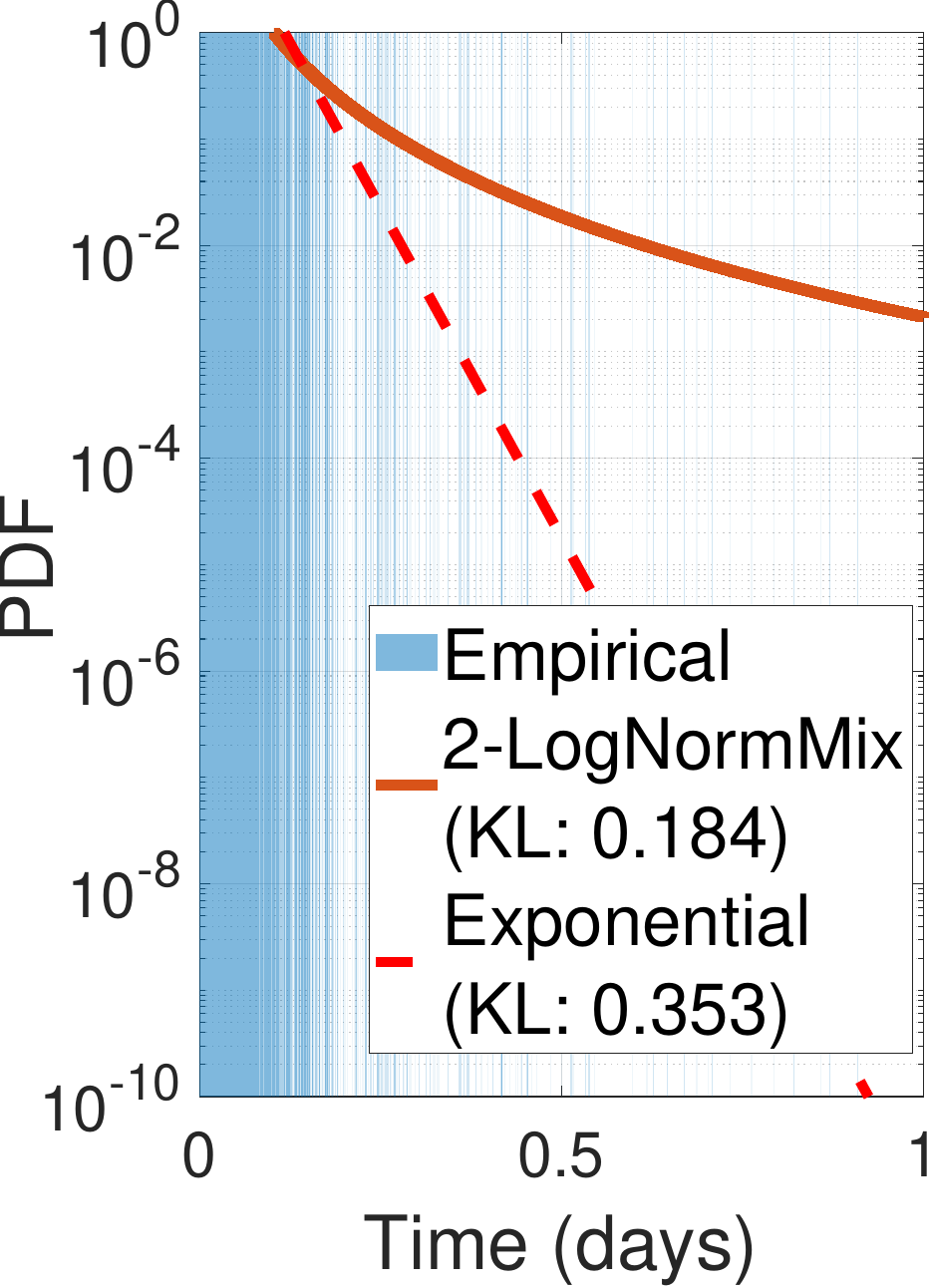}
    \caption{IA fits for Component 1 (Weeks 0–64).}
\end{subfigure}
\hfill
\begin{subfigure}{0.16\textwidth}
    \centering
    \includegraphics[width=\linewidth]{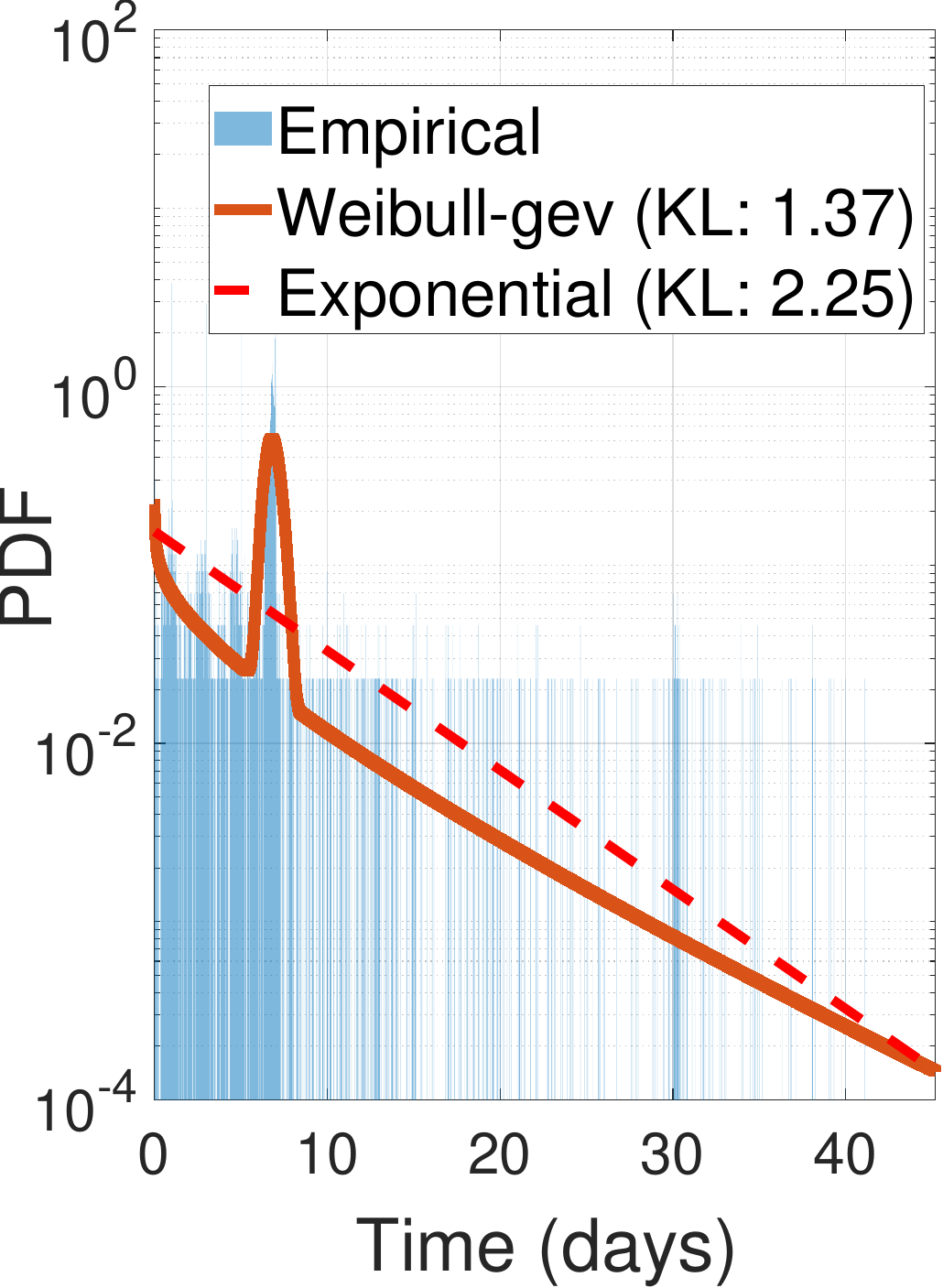}
    \caption{ST fits for Component 1 (Weeks 0–64).}
\end{subfigure}
\hfill
\begin{subfigure}{0.2\textwidth}
    \centering
    \includegraphics[width=\linewidth]{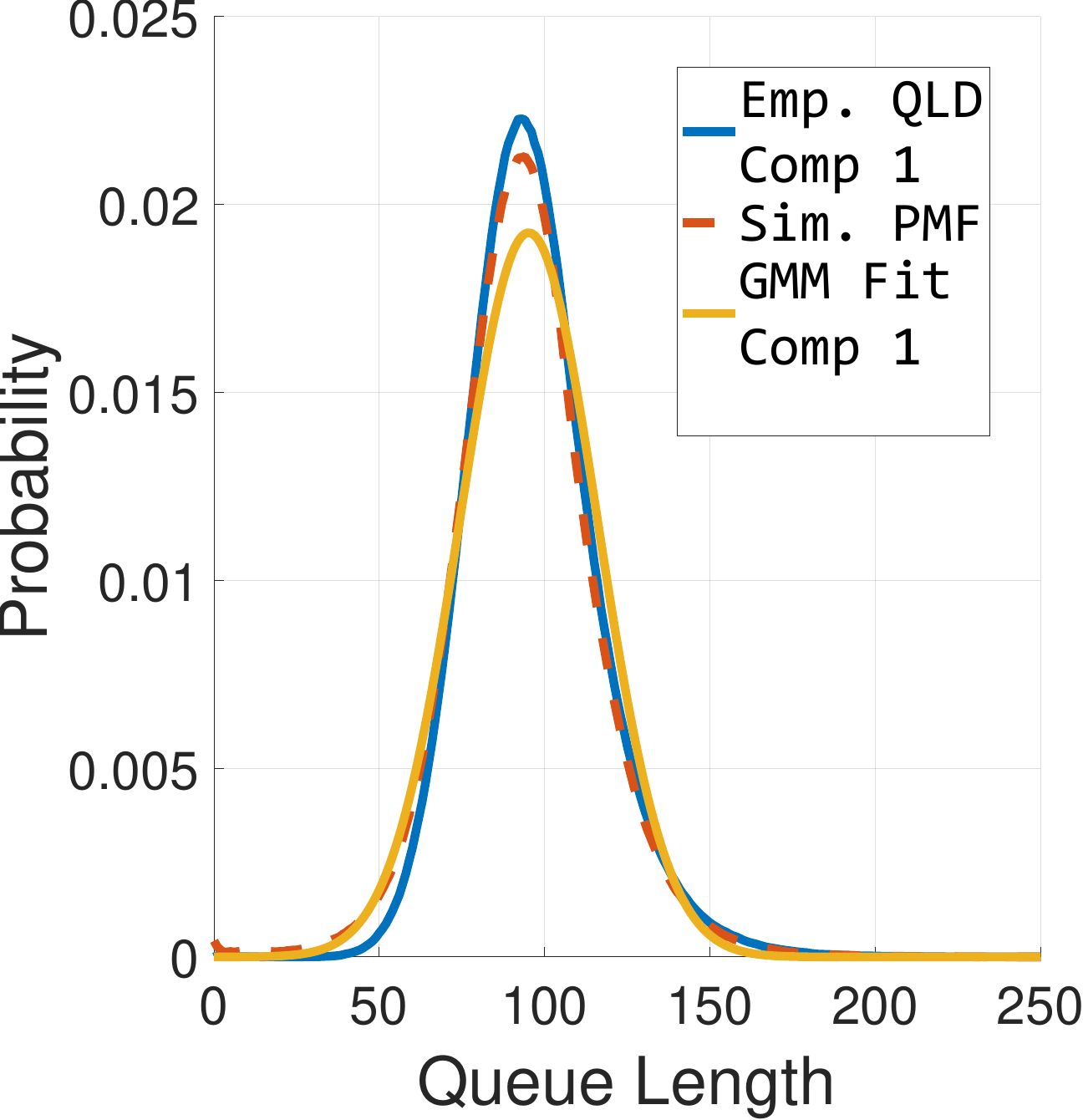}
    \caption{Simulated vs. empirical QLD for Component 1 under fitted parameters.}
\end{subfigure}
\hfill
\begin{subfigure}{0.42\textwidth}
    \centering
    \includegraphics[width=\linewidth]{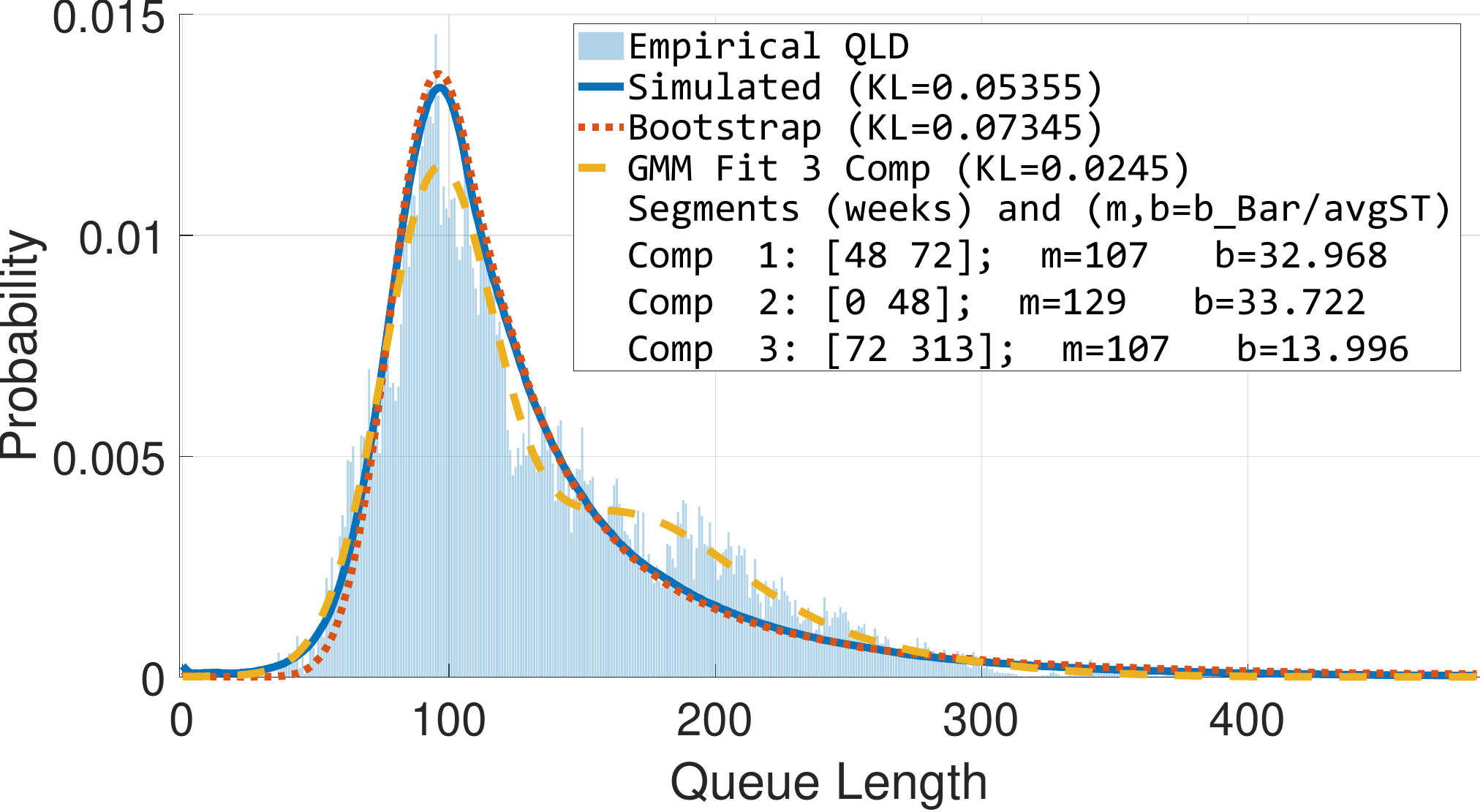}
    \caption{Aggregated empirical QLD with segmented $G/G/m\text{-}b$ simulation, bootstrap estimate, and GMM benchmark.}
\end{subfigure}

\caption{Segment-level fitting and overall QLD validation for the logistics enterprise dataset. The aggregated segmented model closely matches the empirical QLD (KL = 0.05355).}
\label{fig:private_four_row}
\vspace{-1em}
\end{figure*}

Using the segmentation procedure described in Section~\ref{subsec:segmentation_gmm}, we identify three mutually exclusive segments defined over the weekly time axis, with cut points at weeks $[0,48]$, $[48,72]$, and $[72,350]$. These correspond respectively to the calendar intervals Jan.~1,~2019--Oct.~28,~2019, Oct.~28,~2019--Apr.~13,~2020, and Apr.~13,~2020--Feb.~10,~2025. Fig.~\ref{fig:private_fullwidth_three}(a) visualizes these cut points overlaid on the queue-length trajectory. The resulting segmentation isolates two quasi-stationary regimes (Segments~1 and~3), separated by a shorter transition period (Segment~2), demonstrating that the proposed method successfully distinguishes both stable and transitional system dynamics.

Following the identification of quasi-stationary segments, we estimate the queue parameters $(m,b)$ separately for each segment using the fine-grained search procedure described in Section~\ref{subsec:Parameter_Fitting}. Fig.~\ref{fig:private_four_row}(b) provides a representative validation example for the first segment, in which the simulated QLD closely reproduces the empirical distribution under the selected parameter pair.

To obtain necessary capacity parameters, IA and ST should be identified for each segment. Although fine-tuning identifies an effective aggregate service parameter $\bar{b}_i$ for segment~$i$, the average ST varies across segments, and thus $\bar{b}_i$ must be normalized. Specifically, we compute the effective service
capacity as
\( b_i \;=\; \bar{b}_i \,/ \,\mathbb{E}[\mathrm{ST}_i], \)
where $\mathbb{E}[\mathrm{ST}_i]$ is the mean ST in segment~$i$. This normalization ensures that the inferred service capacity reflects the actual throughput of the system rather than segment-specific temporal scaling effects, and is thus essential for consistent parameter interpretation (cf.~Fig.~\ref{fig:model}).

Representative fitted distributions for IA and ST of the first segment are shown (Fig.~\ref{fig:private_four_row}(a)–(b)).  The IA distribution exhibits a heavy-tail, indicating variability in vulnerability discovery rates, with extended periods of low activity interspersed with bursts of arrivals. In contrast, the ST distribution displays a distinct mode around seven days, reflecting patching and service-level targets favor of one week. The presence of this dominant time scale indicates structured patching workflows rather than memoryless service behavior, consistent with Fig.~\ref{fig:temporal_dynamics_singlecol}(b).

Finally, by aggregating the segment-wise simulation results, we obtain the overall QLD shown in Fig.~\ref{fig:private_four_row}(d). The aggregated model accurately reproduces the empirical QLD, yielding a KL of only $0.05355$ between the simulated and empirical QLDs.

Beyond distributional fidelity, the inferred model parameters admit a direct interpretation. In the proposed framework, $m$ corresponds to the number of active personnel resolving tickets (with at most one ticket per worker at a time), while $b$ represents the total capacity measured in jobs per day. The segment-wise analysis reveals that the effective workforce size remains relatively stable at $m \approx 110$, whereas the total capacity decreases over time, from roughly $33$~jobs/day in the early segment to about $14$~jobs/day in the final segment, reflecting measurable changes in throughput.

Independent organizational records enable direct quantitative validation of these estimates as reported in Table~\ref{tab:segment_mb}. The inferred workforce size deviates by less than $1\%$ in the first segment ($m=107$ estimated versus $106$ observed), and remains within $5\%$ across all segments. Similarly, the inferred service capacity in the final segment is $14$~jobs/day, compared to an empirical average of $15.41$~jobs/day, corresponding to a relative error of ~$9\%$. Across segments, the recovered parameters consistently track observed staffing and throughput levels with low relative error.

These results demonstrate that the proposed non-stationary queue reconstruction framework does not merely reproduce QLDs, but accurately identifies the time-varying organizational resources driving backlog dynamics. Importantly, these resource estimates are obtained solely from event-level arrival and closure timestamps, without direct access to staffing data. This ability to recover $m$ and $b$ from traces highlights the practical value of the framework for workforce planning and capacity forecasting in large-scale cyber-security operations.

\begin{table}[t]
\centering
\caption{Segment-wise estimates of the effective number of active defenders ($m$) and aggregate service rate ($b$), with corresponding simulation-derived values ($m_{\mathrm{sim}}, b_{\mathrm{sim}}$).}
\label{tab:segment_mb}
\begin{tabular}{c c c c c c}
\hline
Segment \hspace{-0.1in} &
Weeks \hspace{-0.1in} &
$m$ \hspace{-0.1in} &
$b$ (jobs/day) \hspace{-0.1in} & 
$m_{sim}$ \hspace{-0.1in} &
$b_{sim} (jobs/day)$ \\
\hline
1 \hspace{-0.1in} & $[48,72]$ \hspace{-0.1in}   & 105.99 \hspace{-0.1in} & 23.61 \hspace{-0.1in} & 107 \hspace{-0.1in} & 32.9\\
2 \hspace{-0.1in} & $[0,48]$ \hspace{-0.1in}    & 119.72 \hspace{-0.1in} & 33.10 \hspace{-0.1in} & 129 \hspace{-0.1in} & 33.7\\
3 \hspace{-0.1in} & $[72,313]$ \hspace{-0.1in}  & 103.48 \hspace{-0.1in} & 15.41 \hspace{-0.1in} & 107 \hspace{-0.1in} & 13.9\\
\hline
\end{tabular}
\vspace{-1em}
\end{table}

\section{Conclusion}
In this work, we develop and validate a dynamic queueing-theoretic model to model cyber attack surfaces. The proposed model was evaluated across two distinct and operationally critical environments: software supply chain vulnerability data and large-scale IT service and ticketing systems. A key outcome of this study is the ability to recover latent organizational resources including the workforce size and load per personnel using only event level timestamps, with estimation errors typically within 4–9\%.  Our approach provides a quantitative foundation for operational security planning, enabling security leaders to quantify current staffing, predict resources for a given attack intensity, and evaluate the impacts of shifts in attack frequency or defensive throughput.

Future work could include: 1. applying the framework to actively remediated vulnerabilities to help organizations reduce the 24‑hour activity cycles, 2. integrating real‑time dynamic staffing recommendations, and 3. generalizing the model to unified IT, AI, and human resource infrastructures in tandem.
%; sharing of findings with organizational staffing leadership and enhancing based on feedback, 
\bibliographystyle{IEEEtran}
\bibliography{references_ARXIV}

@article{manadhata2011attackSurface,
  author  = {P. K. Manadhata and J. M. Wing},
  title   = {An Attack Surface Metric},
  journal = {IEEE Transactions on Software Engineering},
  volume  = {37},
  number  = {3},
  pages   = {371--386},
  year    = {2011}
}

@article{harry2025countyAttackSurface,
  author  = {C. Harry and I. Sivan-Sevilla and M. McDermott},
  title   = {Measuring the Size and Severity of the Integrated Cyber Attack Surface Across US County Governments},
  journal = {Journal of Cybersecurity},
  volume  = {11},
  number  = {1},
  pages   = {tyae032},
  year    = {2025}
}

@book{jones2011fair,
  author    = {J. A. Jones},
  title     = {FAIR: Factor Analysis of Information Risk},
  publisher = {Risk Management Insight LLC},
  year      = {2011}
}

@inproceedings{wang2008attackGraphMetric,
  author    = {H. Wang and D. Zhang and S. Jajodia},
  title     = {An Attack-Graph Based Probabilistic Security Metric},
  booktitle = {IFIP Data and Applications Security},
  pages     = {109--124},
  year      = {2008}
}

@article{poolsappasit2012bayesianGraphs,
  author  = {N. Poolsappasit and R. Dewri and I. Ray},
  title   = {Dynamic Security Risk Management Using Bayesian Attack Graphs},
  journal = {IEEE Transactions on Dependable and Secure Computing},
  volume  = {9},
  number  = {1},
  pages   = {61--74},
  year    = {2012}
}

@article{haldar2017vulnEpidemics,
  author  = {K. Haldar and B. K. Mishra},
  title   = {Mathematical Model on Vulnerability Characterization and Its Impact on Network Epidemics},
  journal = {International Journal of System Assurance Engineering and Management},
  volume  = {8},
  number  = {2},
  pages   = {378--392},
  year    = {2017}
}

@inproceedings{feutrill2020queueing,
  author    = {A. Feutrill and M. Roughan and J. Ross and Y. Yarom},
  title     = {A Queueing Solution to Reduce Delay in Processing of Disclosed Vulnerabilities},
  booktitle = {IEEE Conference on Trust, Privacy and Security in Intelligent Systems and Applications (TPS-ISA)},
  pages     = {1--11},
  year      = {2020}
}

@article{khosraviFarmad2020bayesian,
  author  = {M. Khosravi-Farmad and A. Ghaemi-Bafghi},
  title   = {Bayesian Decision Network-Based Security Risk Management Framework},
  journal = {Journal of Network and Systems Management},
  volume  = {28},
  pages   = {1794--1819},
  year    = {2020}
}

@article{ryan2009bayesianThreats,
  author  = {J. J. Ryan and S. D. Dexter},
  title   = {A Bayesian Network Model for Predicting Cyber Security Threats},
  journal = {Journal of Information Assurance and Security},
  volume  = {4},
  number  = {2},
  pages   = {105--114},
  year    = {2009}
}

@article{zhang2018fuzzyBNics,
  author  = {Q. Zhang and C. Zhou and Y.-C. Tian and N. Xiong and Y. Qin and B. Hu},
  title   = {A Fuzzy Probability Bayesian Network Approach for Dynamic Cybersecurity Risk Assessment in Industrial Control Systems},
  journal = {IEEE Transactions on Industrial Informatics},
  volume  = {14},
  number  = {6},
  pages   = {2497--2506},
  year    = {2018}
}

@article{sabur2022graphBasedCloud,
  author  = {A. Sabur and A. Chowdhary and D. Huang and A. Alshamrani},
  title   = {Toward Scalable Graph-Based Security Analysis for Cloud Networks},
  journal = {Computer Networks},
  volume  = {206},
  pages   = {108795},
  year    = {2022}
}

@article{liu2019sequentialAttacks,
  author  = {Q. Liu and L. Xing and C. Zhou},
  title   = {Probabilistic Modeling and Analysis of Sequential Cyber-Attacks},
  journal = {Engineering Reports},
  volume  = {1},
  number  = {4},
  year    = {2019}
}

@article{kotenko2022slrCorrelation,
  author  = {I. Kotenko and D. Gaifulina and I. Zelichenok},
  title   = {Systematic Literature Review of Security Event Correlation Methods},
  journal = {IEEE Access},
  volume  = {10},
  pages   = {43387--43420},
  year    = {2022}
}

@misc{jin2023prometheus,
  author = {X. Jin and others},
  title  = {Prometheus: Infrastructure Security Posture Analysis with AI-Generated Attack Graphs},
  year   = {2023},
  note   = {Preprint}
}

@article{jin2023graphene,
  author  = {X. Jin and others},
  title   = {Graphene: Infrastructure Security Posture Analysis with AI-Generated Attack Graphs},
  journal = {arXiv preprint arXiv:2312.13119},
  year    = {2023}
}

@article{li2024ragAttackGraphs,
  author  = {C. Li and others},
  title   = {Using Retriever-Augmented LLMs to Generate Attack Graphs},
  journal = {arXiv preprint arXiv:2408.05855},
  year    = {2024}
}

@techreport{edgescan2025vulnStats,
  author = {Edgescan},
  title  = {2024 Vulnerability Statistics Report},
  year   = {2025}
}

@article{fang2024llmHack,
  author  = {R. Fang and others},
  title   = {LLM Agents Can Autonomously Hack Websites},
  journal = {arXiv preprint arXiv:2402.06664},
  year    = {2024}
}

@article{mei2024arvo,
  author  = {X. Mei and others},
  title   = {ARVO: Atlas of Reproducible Vulnerabilities for Open Source Software},
  journal = {arXiv preprint arXiv:2408.02153},
  year    = {2024}
}

@misc{berabi2024deepcodeFix,
  author = {B. Berabi and others},
  title  = {DeepCode AI Fix: Fixing Security Vulnerabilities with Large Language Models},
  year   = {2024}
}

@article{yoran2024assistantbench,
  author  = {O. Yoran and others},
  title   = {AssistantBench: Can Web Agents Solve Realistic Tasks on the Open Web?},
  journal = {arXiv preprint arXiv:2407.15711},
  year    = {2024}
}

@inproceedings{liu2024promptInjection,
  author    = {Y. Liu and Y. Jia and R. Geng and J. Jia and N. Z. Gong},
  title     = {Formalizing and Benchmarking Prompt Injection Attacks and Defenses},
  booktitle = {USENIX Security},
  year      = {2024}
}

@book{gautam2012analysis,
  author    = {N. Gautam},
  title     = {Analysis of Queues},
  publisher = {CRC Press},
  year      = {2012}
}

\end{document}